\documentclass[pra,twocolumn,preprintnumbers,amsmath,amssymb]{revtex4}

\usepackage{graphicx}
\usepackage{dcolumn}
\usepackage{bm}
\usepackage{color}
\usepackage{epstopdf}

\usepackage{ulem}

\newcommand{\br}{ {\bm r}}

\def\br{{\bm r}}

\begin{document}



\title{Spatio-temporal thermalization and adiabatic cooling of guided light waves}
\author{Lucas Zanaglia$^1$, Josselin Garnier$^{2}$, Iacopo Carusotto$^3$, Val\'erie Doya$^{1}$, Claire Michel$^{1,4}$, Antonio Picozzi$^{5}$}
\affiliation{$^{1}$ Universit\'e C\^ote d'Azur, CNRS, Institut de Physique de Nice, Nice, France}
\affiliation{$^{2}$ CMAP, CNRS, Ecole polytechnique, Institut Polytechnique de Paris, 91120 Palaiseau, France}
\affiliation{$^{3}$Pitaevskii BEC Center, CNR-INO and Dipartimento di Fisica, Universit\`a di Trento, I-38123 Trento, Italy.}
\affiliation{$^{4}$ Institut Universitaire de France (IUF), 1 rue Descartes, 75005 Paris, France}
\affiliation{$^{5}$ Universit\'e Bourgogne Europe, CNRS, Laboratoire Interdisciplinaire Carnot de Bourgogne ICB UMR 6303, 21000 Dijon, France}


\begin{abstract}
We propose and theoretically characterize three-dimensional spatio-temporal thermalization of a continuous-wave classical light beam propagating along a multi-mode optical waveguide. By combining a non-equilibrium kinetic approach based on the wave turbulence theory and numerical simulations of the field equations, 
we anticipate that thermalizing scattering events are dramatically accelerated by the combination of strong transverse confinement with the continuous nature of the temporal degrees of freedom. 
In connection with the blackbody catastrophe, the thermalization of the classical field in the continuous temporal direction provides a novel intrinsic mechanism for adiabatic cooling and spatial beam condensation. This process of adiabatic cooling is distinct from other mechanisms of thermalization and provides new insights into the dynamics of far-from-equilibrium closed systems and their route to thermalization.
\end{abstract}


\maketitle

{\it Introduction.-}
Fluids of light in propagating geometries are an emerging platform to study the physics of quantum gases. In contrast to the driven-dissipative dynamics of the fluid of light in cavity configurations,  these systems are characterized by a conservative Hamiltonian dynamics~\cite{carusotto22,bloch21,glorieux25}.
In recent years, intense efforts have been devoted to the experimental observation of the 2D relaxation dynamics of monochromatic light towards a fully thermalized equilibrium state~\cite{PRL20,pourbeyram22,mangini22}. In this framework, special attention is being devoted to parabolic (graded-index) multimode optical fiber configurations where spatial confinement prevents beam expansion and ensures efficient nonlinear interactions. These advances were anticipated by spatial beam cleaning experiments~\cite{wright16,krupa17,ferraro23} and related theoretical works~\cite{PRA19,christodoulides19,kottos20,makris22} and have then fueled the emergence of the new field of optical thermodynamics~\cite{christodoulides19,ferraro24,kirsch25,PRL22,podivilov22,kottos24,lu24,PRL23,ferraro25}.

The physics is far richer when one goes beyond the monochromatic assumption and allows for a spatio-temporal (ST) 3D dynamics. In this case, the mapping of light propagation onto a fluid of light can be rigorously formulated at the microscopic quantum level by exchanging the role of the propagation coordinate $z$ and of the physical time $t$, which leads  to a full quantum fluid theory of light~\cite{larre2015propagation}.
In synergy with pioneering experiments on quantum fluctuation features \cite{glorieux22}, the study of 3D fluids of light is closely connected with the strong on-going activity on the spatio-temporal (ST) dynamics of light beams, in particular in multimode waveguides \cite{agrawal,wright22} (multimode solitons \cite{sun24}, conical emission \cite{kibler21},  instabilities \cite{krupa16}, supercontinuum generation \cite{wright16,gentyNC22}).

A natural challenge~\cite{PR14,chiocchetta16,podivilov22} is then to extend optical thermodynamics concepts to the 3D case with the ambitious objective of observing full ST thermalization of light. In waveguide geometries, the 3D extension of the transverse spatial beam-cleaning so far observed only in 2D for monochromatic light would open the intriguing possibility of condensation also in the spectral domain. 

In this Letter, we take inspiration from on-going efforts in the cold-atom context~\cite{Mandonnet2000,DGO_th,DGO_exp,Raithel,Schreck} to propose a strategy to observe an efficient 3D thermalization of classical waves in both the spatial and the temporal directions. While spatial confinement is guaranteed by a multimode waveguide geometry, the use of a temporally continuous incoherent input wave, instead of the usual coherent pulses~\cite{PRL20,pourbeyram22,mangini22,podivilov22,PRL22,ferraro24,kirsch25,PRL23,ferraro25,wright16,krupa17,ferraro23}, guarantees that the beam remains homogeneous in the temporal direction with no need for confinement to overcome pulse broadening effects. 
In contrast to the extremely slow thermalization of quantum light predicted in~\cite{chiocchetta16}, bosonic stimulation effects~\cite{davis01,PRL05,nazarenko11,blakie08,krstulovic11} are anticipated to strongly accelerate thermalization of classical waves. As a key feature of ST thermalization, we show how the presence of a continuum of temporal modes overcomes the obstacles imposed by the discrete waveguide modes onto the spatial-only thermalization~\cite{nazarenko11,lvov10,onorato23}. On top of a fast convergence toward a local Rayleigh-Jeans (RJ) ST equilibrium, we anticipate an unexpected process of adiabatic cooling, which opens new perspectives on the non-equilibrium dynamics of closed systems and on their paths to thermalization~\cite{nazarenko23,Gasenzer23,erne18,prufer18,moreno25,gazo25}.

\begin{center}
\begin{figure}
\includegraphics[width=1\columnwidth]{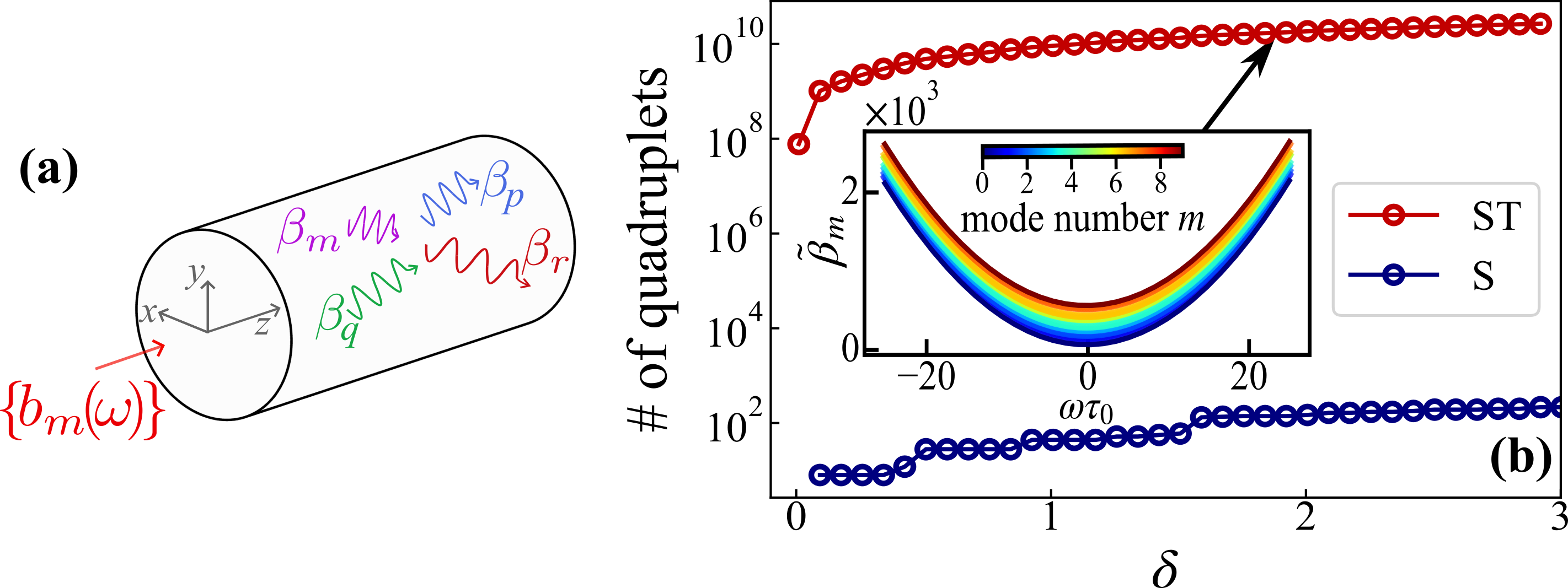}
\caption{
\baselineskip 10pt
{\bf Quasi-resonances.} 
(a) Schematic visualization of the waveguide. (b) Number of mode quadruplets $\{m,p,q,r\}$ verifying a resonance condition  in the spatial-only $|\Delta \beta^S_{mpqr}| L_{\rm nl} < \delta$ (blue line) and in the ST $|\Delta \beta^{ST}_{mpqr}| L_{\rm nl} < \delta$ (red line) cases.
The temporal degree of freedom leads to a dramatic enhancement of the number of quasi-resonances.
Inset: Modal dispersion relations ${\tilde \beta}_m(\omega)$ (in mm$^{-1}$) used to compute the resonant quadruplets in the ST case.
Parameters: $L_{\rm nl}=0.3$m, $\tau_0=\sqrt{|\kappa_2| L_{\rm nl}}$, 
${\omega}_c=25/\tau_0$ (the waveguide configuration is described in the Appendix, Sec.~IV).
}
\label{fig:1} 
\end{figure}
\end{center}

\medskip
\noindent
{\it The NLSE model.}
Consider a random classical field with central frequency $\tilde{\omega}_o$, propagating along the $z$-axis of a waveguide, see Fig.~\ref{fig:1}(a).
At a simplest yet accurate level of approximation, we describe the 
slowly varying field envelope $\psi(t,\br,z)$ around the carrier frequency $\tilde{\omega}_o$ and wavevector $k_o$ in terms of a nonlinear Schr\"odinger equation (NLSE)~\cite{larre2015propagation,glorieux25}:
\begin{align}
i\partial_z \psi = -\frac{1}{2k_o} \nabla_\perp^2 \psi + V(\br) \psi + \kappa_2 \partial_t^2 \psi - \gamma_o |\psi|^2\psi.
\label{eq:nls_rt}
\end{align}
The propagation distance $z$ plays the role of an evolution `time' variable, while $t$ plays the role of a spatial variable in the reference frame that propagates with group-velocity $v_g^{-1}=\partial_{\tilde \omega} k({\tilde \omega}_o)$. Here, $k({\tilde \omega})$ is the dispersion relation as a function of the frequency ${\tilde \omega}$ and the carrier satisfies $k_o=k({\tilde \omega}_o)$. The parameter $\kappa_2=\frac{1}{2} \partial_{\tilde \omega}^2 k({\tilde \omega}_o)$ denotes the dispersion coefficient, $\gamma_o$ the nonlinear coefficient, and the transverse trapping potential $V(\br)$ accounts for the refractive index profile in the transverse $\br = (x,y)$ directions.

We introduce a real-valued orthonormal basis $u_m(\br)$, with eigenvalues $\beta_m$, solutions of the transverse waveguide problem $\beta_m u_m(\br)=[-(2k_o)^{-1} \nabla_\perp^2 + V(\br)] u_m(\br)$. We then expand the field $\psi(t,\br,z)$ 
in terms of its modal amplitudes $ b_m({\omega},z)  = \iint \!dt\, d\br\, \psi(t,\br,z) \exp( i {\omega} t) \, u_m(\br)$, which are governed by (see Appendix, Sec.~I):
\begin{align}
i \partial_z {b}_m = {\tilde \beta}_m({\omega})   {b}_m  -  \gamma \Gamma_m({\omega})  P_{m}({\bm b})\, ,
\label{eq:upe}
\end{align}
where the linear and nonlinear dispersion relations respectively read ${\tilde \beta}_m({\omega})=\beta_m - \kappa_2 {\omega}^2$ and $\Gamma_m({\omega})=1$, with ${\omega}={\tilde \omega}-{\tilde \omega}_o \ll {\tilde \omega}_o$ the frequency offset from the carrier ${\tilde \omega}_o$, and $\gamma=\gamma_o/(2\pi)^2$. The interaction among modes mediated by the optical nonlinearity is  
$P_{m}({\bm b})=\sum_{pqr} W_{mpqr} \int {b}_p({\omega}_1,z) {b}_q^*({\omega}_2,z){b}_r({\omega}_3,z) \delta_{\omega} d{\omega}_1 d{\omega}_2 d{\omega}_3$, where  $\delta_{\omega}=\delta({\omega}-{\omega}_1+{\omega}_2-{\omega}_3)$ and $W_{mpqr} = \int u_m(\br)u_p(\br)u_q(\br)u_r(\br) d\br$ accounts for the spatial overlap among the eigenmodes. 
In realistic configurations, the truncated potential $V(\br)$ guides a finite number of modes, labeled by the indices $m,p,q,r=0,..,M-1$, where $M$ is the total number of propagative modes supported by the waveguide. 
Note that $\beta_m$ and $W_{mpqr}$ do not depend on the frequency $\omega$, because we have expanded the field on $\omega$-independent eigenmodes $\{u_m(\br)\}$, see Appendix (Sec.~I).

While the NLSE is able to capture the essential physics in a simple and broadly accessible form, a more accurate theory -- based on the unidirectional propagation equation (UPE)~\cite{kolesik02,couairon07,kolesik11,moloney16,bejot19} -- is presented in the Appendix (Sec.~I). In a waveguide geometry, the UPE for the mode amplitudes $b_m(\omega,z)$ has the form Eq.(\ref{eq:upe}) and the generalized 
linear and nonlinear dispersion effects are described by suitable functions ${\tilde \beta}_m({\omega})$ and $\Gamma_m({\omega})$ (Appendix, Sec.~I).

\medskip
\noindent
{\it Role of resonances in the thermalization process.}
Non-equilibrium light thermalization is crucially driven by interactions between modes mediated by optical nonlinearities. For relatively weak nonlinearities, the efficiency of these processes strongly relies on the presence of resonances among quartets of modes. The difference in behavior between the spatial-only case for monochromatic light and the ST case is illustrated in Fig.~\ref{fig:1}(b), for the concrete example of a step-index waveguide supporting $M=10$ modes. In particular, we show the number of mode quadruplets $\{m,p,q,r\}$ that satisfy a quasi-resonance condition $|\Delta \beta_{mpqr}| L_{\rm nl} < \delta$ in the two cases. Here $L_{\rm nl} \simeq 1/(|\gamma_o| {\bar N})$ is the nonlinear length and ${\bar N}$ the average intensity.

In the spatial case, typically large values 
of the detunings $\Delta \beta^S_{mpqr} =  \beta_m+\beta_q-\beta_p-\beta_r$ suppress thermalization in generic waveguides~\cite{PRA19}, unless resonances are enforced by symmetry reasons, e.g., in parabolic-shaped multimode fibers~\cite{ferraro23}.
In the ST case, instead, $\beta_m$ is replaced by the continuous frequency-dependent dispersion ${\tilde \beta}_m(\omega)=\beta_m-\kappa_2 \omega^2$. The inset of  Fig.~\ref{fig:1}(b) reports an example of ${\tilde \beta}_m(\omega)$ 
in the anomalous dispersion regime ($\kappa_2<0$). The four-mode detuning parameter now reads $\Delta {\tilde \beta}_{mpqr}^{\omega 123}={\tilde \beta}_m(\omega)-{\tilde \beta}_p(\omega_1)+{\tilde \beta}_q(\omega_2)-{\tilde \beta}_r(\omega_3)$: The extra degree of freedom introduced by $\omega$ 
ensures the existence of efficient resonances $|\Delta \beta^{ST}_{mpqr}| L_{\rm nl} \ll 1$ for the different quartets of modes, where for each quartet, $\Delta {\beta}^{ST}_{mpqr}={\rm min}(\Delta {\tilde \beta}_{mpqr}^{\omega 123})$ is optimized 
over the frequencies $\{ \omega, \omega_{1,2,3} \}$ in the range $|\omega| \le \omega_c$ of our simulations.
As evidenced in Fig.~\ref{fig:1}(b), the number of quasi-resonances in the ST case exceeds the one in the spatial-only case by several orders of magnitude, guaranteeing efficient thermalization.
This result is general and totally independent of the specific waveguide configuration considered.

\medskip
\noindent
{\it Spatio-temporal wave turbulence.}
The intuitive physical picture about resonant interactions depicted in Fig.~\ref{fig:1} can be formalized in the framework of the wave turbulence theory, which provides a detailed nonequilibrium description of the irreversible thermalization process \cite{zakharov92,Newell01,nazarenko11,Newell-Rumpf,shrira-nazarenko13,babin07,churkin15,laurie12,onorato23,PR14}. 
As anticipated in Fig.~\ref{fig:1}, in the spatial case the transverse confinement of the field imposed by the potential $V(\br)$ leads to a {\it discrete} set of resonant interactions, which can suppress efficient quasi-resonances and ultimately freezes the thermalization process. 
This aspect was discussed in the context of {\it discrete} wave turbulence   \cite{nazarenko11,lvov10,onorato23,wang20,cherroret21,shepelyansky23,kottos23}.  
In this discrete turbulence regime, 
it is not possible to derive a {\it continuous} kinetic equation describing spatial-only thermalization, see Sec.~III in the Appendix.

Conversely, we show that the continuous nature of the temporal degrees of freedom restores efficient ST resonances, enabling the derivation of a {\it hybrid discrete-continuous} ST kinetic equation, 
involving discrete sums over the spatial modes and continuous integrals over the temporal spectrum. 
We consider random initial conditions with statistically stationary (in time) distribution.
The initial spectrum $n_m(\omega,0)$ is such that $\left< {b}_m(\omega, 0) {b}_{p}^*(\omega', 0) \right> ={n}_m(\omega, 0) \delta_{mp} \delta(\omega-\omega')$,
where $\delta_{mp}$ is the Kronecker symbol
and the brackets $\left< \cdot \right>$ denote an average with respect to the distribution of the random initial conditions.
In the weakly nonlinear regime, 
we derive in the Appendix (Sec.~II) the kinetic equation for the evolution of the ST spectrum $n_m(\omega,z)$ such that $\left< {b}_m(\omega, z) {b}_{p}^*(\omega', z) \right> ={n}_m(\omega, z) \delta_{mp} \delta(\omega-\omega')$:
\begin{align}
\partial_z {n}_m(\omega,z)=4\pi \gamma^2 \sum_{pqr} \int d\omega_{1-3} |W_{mpqr}|^2 {\bm M}_{mpqr}({\bm {n}}) 
\nonumber \\ 
\quad \times \delta(\omega-\omega_1+\omega_2-\omega_3) \delta(\Delta {\tilde \beta}_{mpqr}^{\omega 123}) 
\label{eq:kin}
\end{align}
with the cubic nonlinear term ${\bm M}_{mpqr}({\bm {n}})={n}_p(\omega_1) {n}_q(\omega_2) {n}_r(\omega_3)+{n}_m(\omega) {n}_p(\omega_1)  {n}_r(\omega_3)-{n}_m(\omega)  {n}_q(\omega_2) {n}_r(\omega_3)-{n}_m(\omega) {n}_p(\omega_1) {n}_q(\omega_2)$. 
Note that degenerate modes have been neglected in the derivation, see the Appendix.

The kinetic Eq.(\ref{eq:kin}) conserves the particle number $N = \sum_m  \frac{1}{(2\pi)^2} \int {n}_m(\omega) d\omega$, the momentum $P = \sum_m \frac{1}{(2\pi)^2}  \int \omega {n}_m(\omega) d\omega$, and the kinetic energy  
$E = \sum_m  \frac{1}{(2\pi)^2}\int {\tilde \beta}_m(\omega) {n}_m(\omega) d\omega$
~\footnote{Note that the factor $1/(2\pi)^2$ comes from the definition of the particle number for a statistically stationary in time field solution of the NLSE, $N=\int \left<  |\psi(t,\br,z)|^2\right> d\br$.}.
At the same time, it exhibits an $H-$theorem of entropy growth $\partial_z S(z) \ge 0$ for the nonequilibrium entropy $S(z)=\sum_m  \int \log\big({n}_m(\omega)\big) d\omega$: At variance with the NLSE (\ref{eq:upe}) that is formally reversible (Hamiltonian) in `time' $z$, it 
then  describes the actual nonequilibrium process of ST thermalization toward the RJ equilibrium:
\begin{equation}
n_m^{RJ}(\omega)
= \frac{T }{{\tilde \beta}_m(\omega){-\lambda {\omega}}-\mu} ,
\label{eq:n_rj_general}
\end{equation}
where the temperature $T$, chemical potential $\mu$, and average `velocity' $\lambda$, are determined from the three conserved quantities 
($N, P, E$).
Note that, in the framework of the UPE, the generalized linear and nonlinear dispersive effects enrich the form of the RJ distribution, yielding $n_m^{RJ}(\omega)= T  \Gamma_m(\omega)/\big({\tilde \beta}_m(\omega)-\lambda {\omega}-\mu\big)$, see Figs.~6-7 in the Appendix.

\begin{center}
\begin{figure}[t]
\includegraphics[width=1\columnwidth]{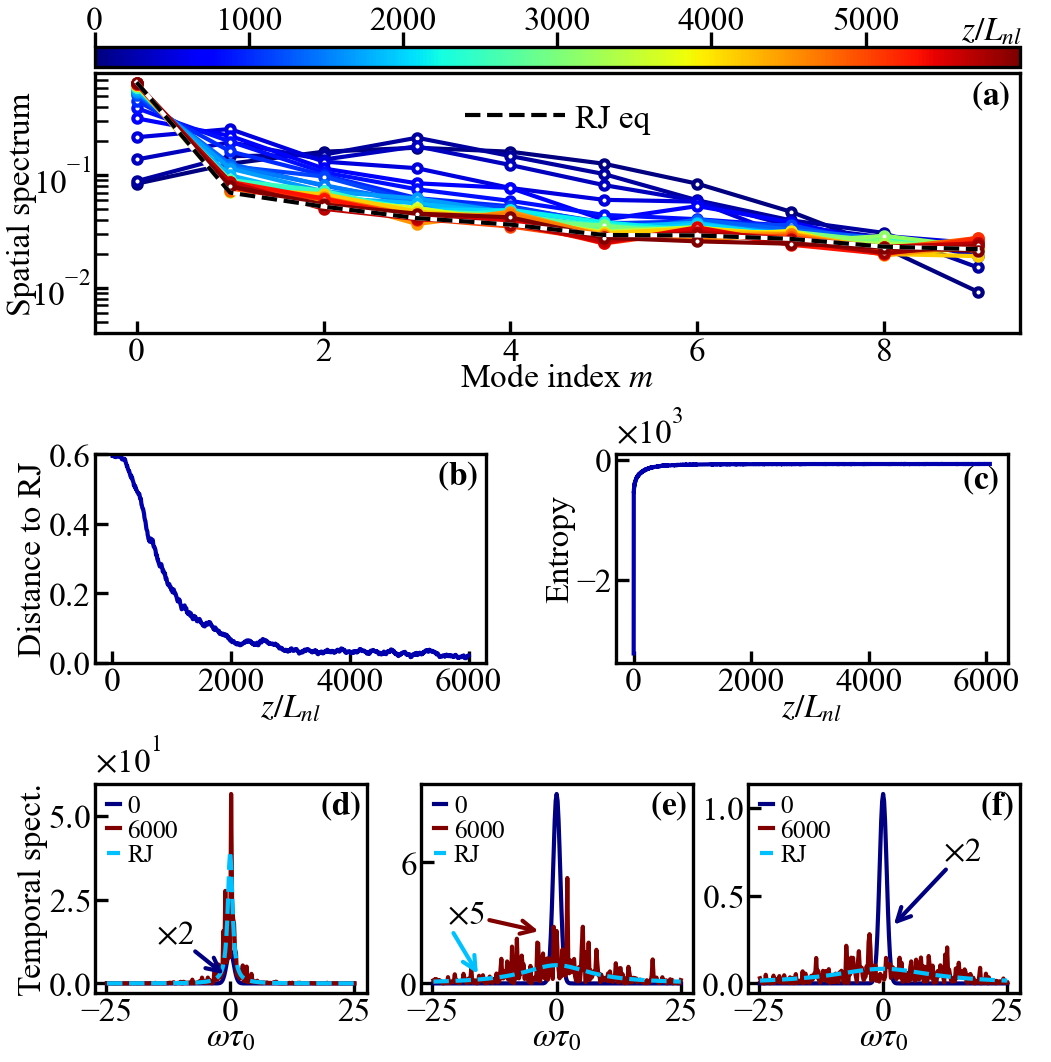}
\caption{
\baselineskip 10pt
{\bf Spatio-temporal thermalization.} 
(a) Simulation of NLSE~(\ref{eq:upe}): Evolution of spatial modal occupation $N_m^{ST}/N$, showing the relaxation to the equilibrium RJ distribution Eq.(\ref{eq:n_rj_general}) (dashed black line).
(b) Evolution of the distance ${\cal D}^{ST}(z)$ to equilibrium, whose decrease to zero evidences ST thermalization. 
(c) This irreversible process is  characterized by a monotonous growth of entropy, as described by the $H-$theorem of the wave turbulence kinetic Eq.(\ref{eq:kin}).
The distance and entropy evolutions are in contrast with those of the spatial case, see Figs.~\ref{fig:3}(b)-(c).
Temporal spectrum $|b_m(\omega, z)|^2$  of the fundamental mode $m=0$ (d), intermediate mode $m=5$ (e), highest mode ($m=9$) (f), at $z=0$ (dark blue) and $z = 5000 L_{\rm nl}$ (red), showing thermalization to RJ spectra (dashed light blue).
Parameters: step-index waveguide supporting $M=10$ modes (see Fig.~\ref{fig:1}(b)), with anomalous dispersion and defocusing nonlinearity, $\tau_0=\sqrt{|\kappa_2| L_{\rm nl}}$, $L_{\rm nl}=0.3$m, 
${\omega}_c=25/\tau_0$, $\sigma_\omega=1/\tau_0$, see the Appendix.
}
\label{fig:2} 
\end{figure}
\end{center}

\medskip
\noindent
{\it Spatio-temporal simulations.}
To put these ideas on quantitative grounds, 
we have performed simulations of the NLSE~(\ref{eq:upe}).
The initial condition is a spatio-temporally incoherent field: the different $(\omega,m)$-components $b_m(\omega,z=0)$ are independent complex-valued Gaussian random variables of zero mean; each spatial mode $m$ has a Gaussian spectrum $\sim \exp(-\omega^2/\sigma_\omega^2)$, with the same width $\sigma_\omega=1/\tau_0$ for all modes, and different amplitudes, see dark blue curves in Fig.~\ref{fig:2}(a,d-f).

ST thermalization to the RJ equilibrium Eq.(\ref{eq:n_rj_general}) is illustrated in Fig.~\ref{fig:2}. Here, the parameters $(T, \lambda, \mu)$ in Eq.(\ref{eq:n_rj_general}) are computed by considering the frequency cutoff ${\omega}_c \tau_0=25$  of the spectral grid used in the simulation, see Sec.~IV in the Appendix. In order to compare the ST dynamics with the spatial-only dynamics (see Fig.~\ref{fig:3} below), we report in Fig.~\ref{fig:2}(a) the evolution of the spatial mode distribution by integrating over the temporal frequencies, $N_m^{ST}(z)=\frac{1}{(2\pi)^2}\int n_m(\omega,z) d\omega$.
ST thermalization is confirmed by the evolution of the distance to RJ equilibrium ${\cal D}^{ST}(z)=\sum_m |N_m^{ST}(z) - N_m^{RJ}|  / \sum_m (N_m^{ST}(z) + N_m^{RJ})$,  with $N_m^{RJ}=\frac{1}{(2\pi)^2}\int n_m^{RJ}(\omega) d\omega$ (note that ${\cal D}^{ST}$ is bounded, $0 \le {\cal D}^{ST} \le 1$). As evidenced in Fig.~\ref{fig:2}(b), the distance ${\cal D}^{ST}(z)$ decreases to zero during propagation, while the temporal spectra converge to those predicted by the RJ equilibrium, see Fig.~\ref{fig:2}(d-f). Note that, to avoid the formation of temporal solitons and the subsequent freezing of the thermalization process, a defocusing nonlinearity had to be used in the anomalous dispersion regime considered in Fig.~\ref{fig:2}. Of course, temporal solitons could be avoided also in the focusing regime by considering the normal dispersion regime.

\begin{center}
\begin{figure}[t]
\includegraphics[width=1\columnwidth]{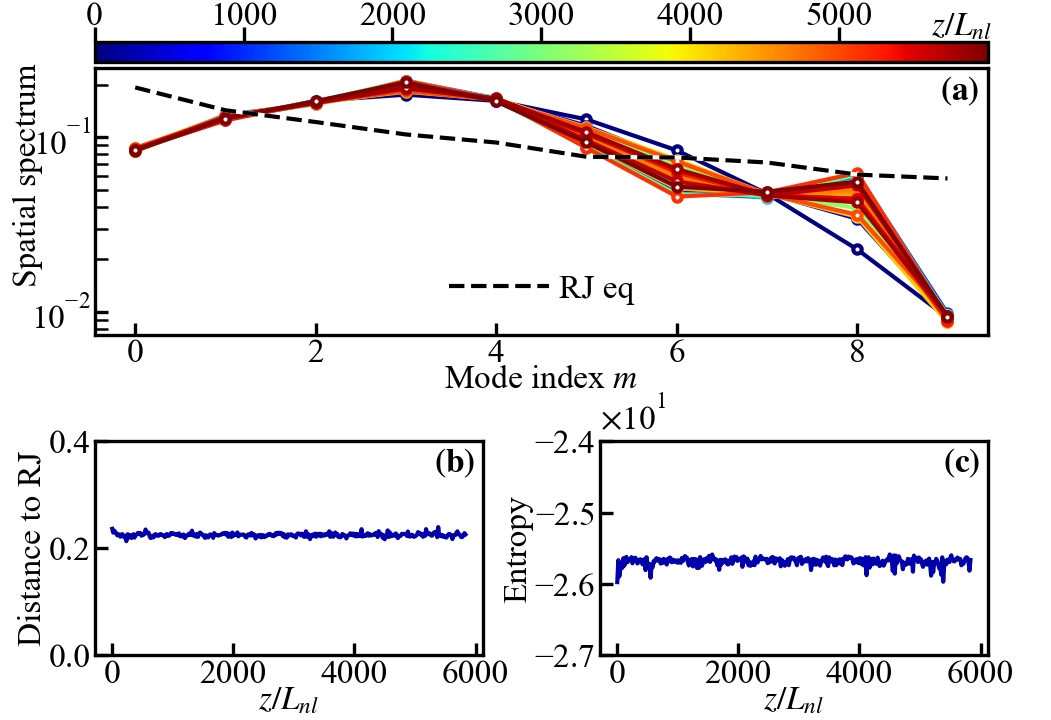}
\caption{
\baselineskip 10pt
{\bf Pure spatial dynamics: Frozen thermalization.} 
(a) Simulation of Eq.(\ref{eq:nls}): 
Evolution of the spatial spectrum $N_m^{S}/N$
starting from the same initial condition as in the ST simulation in Fig.~\ref{fig:2}.
The thermalization process is frozen, as evidenced by the distance ${\cal D}^{S}(z)$ to RJ equilibrium (b), and the entropy (c), which, in contrast to the ST case of Fig.~\ref{fig:2}(b)-(c), maintain a constant value at long times. Because of the large fluctuations of individual realizations, an average has been taken over 14 realizations.
}
\label{fig:3} 
\end{figure}
\end{center}

\medskip
\noindent
{\it Spatial vs spatio-temporal dynamics.}
To clearly evidence the key role of the temporal degrees of freedom, we compare the ST simulation in Fig.~\ref{fig:2}, with the equivalent simulation in the purely spatial problem,  
where the spatial modal amplitudes $b_m^S(z)$ are ruled by 
\begin{align}
i \partial_z {b}_m^S = \beta_m {b}_m^S  -  \gamma_o \sum_{p,q,r} W_{mpqr}{b}_p^S {b}_q^{S*} {b}_r^S.
\label{eq:nls}
\end{align}
The evolution of the spatial mode distribution $N_m^{S}(z)=|b_m^{S}(z)|^2$ is then compared to the ST evolution $N_m^{ST}(z)$, considering the same initial condition, 
$N_m^{S}(z=0)=N_m^{ST}(z=0)$.
The comparison is striking: 
In the ST case  (Fig.~\ref{fig:2}) the evolution exhibits a fast relaxation to equilibrium, whereas in the spatial case (Fig.~\ref{fig:3})  the thermalization process is frozen.
This is confirmed by the evolution of the distance ${\cal D}^{S}(z)=\sum_m |N_m^{S}(z) - N_m^{RJ}|  / \sum_m (N_m^{S}(z) + N_m^{RJ})$ to the spatial RJ equilibrium $N_m^{RJ}=T^{S}/(\beta_m-\mu^{S})$: In contrast to ${\cal D}^{ST}(z)$ in Fig.~\ref{fig:2}(b),  ${\cal D}^{S}(z)$ does not decrease in Fig.~\ref{fig:3}(b).


Further insight on the profound distinction between the ST dynamics and the pure spatial dynamics is offered by the evolution of entropy. In the ST case, the irreversible process of thermalization is featured by a monotonic increase of entropy, as dictated by the $H-$theorem of entropy growth inherent to the kinetic Eq.(\ref{eq:kin}), see Fig.~\ref{fig:2}(c).
Note that the entropy growth saturates rapidly compared to the slow decrease of the distance ${\cal D}^{ST}$, since this latter is mainly driven by highly populated low-order modes, while the entropy is more  sensitive to weakly populated high-order modes. 
On the other hand, the spatial dynamics shown in Fig.~\ref{fig:3}(c) evolves in a discrete wave turbulence regime governed by a {\it reversible} system of kinetic equations: this system 
does not exhibit an $H-$theorem of entropy growth and explains the frozen thermalization observed in Fig.~\ref{fig:3}, 
see Fig.~5 in the Appendix. 
\begin{center}
\begin{figure}[t]
\includegraphics[width=1\columnwidth]{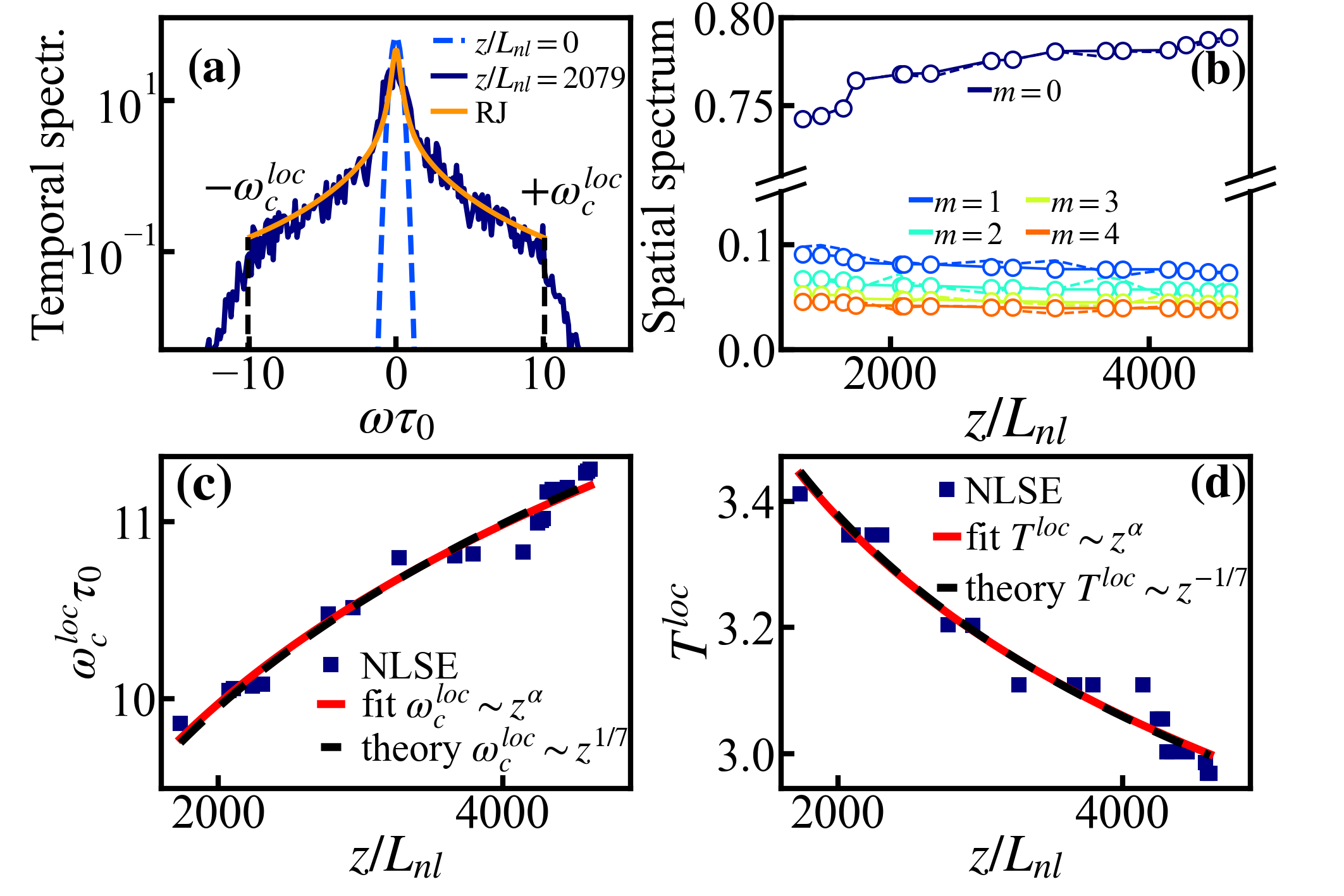}
\caption{
\baselineskip 10pt
{\bf Local-equilibrium route to ST thermalization and adiabatic cooling.} 
(a) Mode-integrated temporal spectrum of the field $\sum_m |b_m(\omega,z)|^2$ at $z=0$ (light blue), $z=2079\,L_{\rm nl}$ (dark blue), and local RJ equilibrium distribution over the reduced frequency window $[-{\omega}_{c}^{loc}(z),{\omega}_{c}^{loc}(z)]$ (orange).
(b) Modal population $N_m^{loc}/N$ computed from the local RJ equilibrium (circles), and modal population $N_m^{ST}/N$ in NLSE simulation (solid lines). 
Corresponding evolutions during propagation of local frequency cut-off $\omega_c^{loc}(z)$ (c), and local temperature $T^{loc}(z)$ (d): Results of NLSE simulations (squares) are fitted by a power-law $\sim z^\alpha$ (red lines), showing quantitative agreement with the theory Eq.(\ref{eq:scaling}) (dashed black lines).  
The decrease in $T^{loc}(z)$ reflects an adiabatic cooling, which drives a spatial beam condensation characterized by the growth of $N_0^{loc}(z)/N$ in (b).
Parameters: step-index waveguide supporting $M = 5$ modes, with anomalous dispersion and defocusing nonlinearity, $L_{\rm nl}=0.06$m, ${\omega}_c=25/\tau_0$, $\sigma_\omega = 0.4/\tau_0$.
}
\label{fig:4} 
\end{figure}
\end{center}

\medskip
\noindent
{\it Adiabatic cooling.}
A well-known issue of classical field theories is the occurrence of ultraviolet (UV) divergences in the thermal equilibrium state, the so-called black-body catastrophe~\cite{zakharov92,nazarenko11,blakie08}. In our configuration, this issue is naturally tamed in the transverse direction by the finite number of modes of the waveguide, but gives rise to a rich physics in the temporal direction. 
Starting from the very non-thermal initial state with a short-tailed Gaussian distribution considered in our simulations, after a transient, 
the optical field approaches at each $z$ a local quasi-equilibrium state that closely approximates a truncated RJ equilibrium distribution within a limited RJ window $[-{\omega}_{c}^{loc}(z),{\omega}_{c}^{loc}(z)]$ and quickly drops to zero outside it, as illustrated in the mode-integrated spectrum shown in Fig.~\ref{fig:4}(a). 
The accuracy of the process of local thermalization is evidenced by the remarkable agreement visible in Fig.~\ref{fig:4}(b) between the modal populations $N_m^{loc}(z)$ predicted by the local RJ equilibrium (circles), and the ones $N_m^{ST}(z)$ computed from the NLSE simulation (solid lines).

For increasing $z$, the bounded window  
$[-{\omega}_{c}^{loc}(z),{\omega}_{c}^{loc}(z)]$ of the local RJ equilibrium expands, as shown in Fig.~\ref{fig:4}(c). Correspondingly, the local thermodynamic parameters also display a marked $z$ dependence, as expected by imposing conservation of $(N,P,E)$ in the presence of the $z$-dependent 
truncation at ${\omega}_{c}^{loc}(z)$. 
The local temperature $T^{loc}(z)$ plotted in Fig.~\ref{fig:4}(d) displays a monotonic decrease due to the continuous transfer (at a constant energy $E=$const) of the incoherent beam fluctuations into the high-energy tails of the spectrum distribution. 
The detailed evolution of  
${\omega}_{c}^{loc}(z)$ and 
$T^{loc}(z)$ depend on all physical parameters $\gamma$, $\kappa_2$ and $W_{mpqs}$ that appear  in the kinetic Eq.(\ref{eq:kin}). It is possible, however, to derive from Eq.(\ref{eq:kin}) the following theoretical scaling law that relates ${\omega}_{c}^{loc}(z)$ and $T^{loc}(z)$ to the propagation distance $z$ and that is valid when 
$z \gg L_{\rm nl}$ (see Sec.~V in the Appendix):
\begin{align}
T^{loc}(z) \sim 1/\omega_c^{loc}(z) \sim 1/z^{1/7}.
\label{eq:scaling}
\end{align}
As illustrated in Fig.~\ref{fig:4}(c)-(d), this power-law behavior is confirmed by NLSE simulations, whose results are fitted by power laws $\sim z^\alpha$, with exponents $\alpha=0.141$ for $\omega_c^{loc}(z)$, and $\alpha=-0.139$ for $T^{loc}(z)$, in remarkable agreement with the value $1/7\sim 0.143$ in Eq.(\ref{eq:scaling}).
Note that, to better highlight the asymptotic behavior Eq.(\ref{eq:scaling}) in Fig.~\ref{fig:4}, the nonlinear strength was increased, while the number of modes and the initial spectral widths were reduced with respect to Fig.~\ref{fig:2} -- we checked that all our predictions on the adiabatic cooling are physical and do not depend on the numerical spectral grid, see the Appendix.

As a direct consequence of this {\it conservative adiabatic cooling}, a marked beam cleaning is visible in Fig.~\ref{fig:4}(b) together with a transverse condensation effect where the population gets macroscopically concentrated in the fundamental waveguide mode, $N_0^{ST} \gg N_{m \neq 0}^{ST}$.
Most interestingly, while in the numerical simulations shown in this Letter the efficiency of the cooling process is artificially limited by the numerical cut-off ($\omega_c$) imposed by the simulation (see Sec.~IV in the Appendix), in a real physical system the UV-divergent value of the energy stored in the tails of the classical distribution would, in principle, allow for an arbitrary large decrease of $T^{loc}(z)$. 
This conclusion, however, holds only within the validity range of our classical, conservative model equation, which neglects material absorption or higher-order dissipative nonlinearities. In addition, at large frequencies the classical model ultimately breaks down, as quantum statistical effects become relevant and drive the spectrum toward a Bose–Einstein distribution~\cite{chiocchetta16,davis01,blakie08}, thereby imposing an ultimate limitation to adiabatic cooling. 
We anyway anticipate that, given the small value of typical nonlinear media, this is expected to occur at extremely large propagation lengths $z$~\cite{chiocchetta16}.

\medskip
\noindent
{\it Conclusions and outlook.}
We developed a general wave turbulence framework for ST thermalization of light waves in multimode Kerr waveguides. For the sake of clarity, the theory and simulations presented in the main text are carried out within a NLSE formalism, but they are extended to more general and accurate UPE~(\ref{eq:upe}) in the Appendix (Sec.~I). 
In contrast to the frozen spatial thermalization of monochromatic light, we find that the presence of the continuous temporal degrees enable efficient ST thermalization. 
As a consequence of the blackbody catastrophe of classical fields, our route to ST thermalization unveils an intrinsic adiabatic cooling mechanism, whereby the field fluctuations are transferred to high-frequency components along the time dimension, enabling a virtually unlimited spatial beam-cleaning condensation. 
This adiabatic cooling is inherently conservative and therefore is to be contrasted to conventional evaporative cooling techniques in Bose-Einstein condensates and to the self-similar processes of (pre-)thermalization~\cite{nazarenko23,Gasenzer23,erne18,prufer18,moreno25,gazo25}. As such, it provides valuable insight into the nonequilibrium dynamics of Hamiltonian systems and their pathways to thermalization. 
More broadly, our conservative wave-guided light configuration could enable full 3D condensation in the quantum regime, opening new avenues for ST beam cleaning and coherent light generation.

\medskip
\noindent
{\it Acknowledgments.} Funding was provided by Agence Nationale de la Recherche (Grants No. ANR-23-CE30-0021, No. ANR-21-ESRE-0040). Calculations were performed using HPC resources from Universit\'e C\^ote d’Azur’s Center Azzura, and DNUM-CCUB (Universit\'e de Bourgogne Europe). I.C. acknowledges financial support by the PE0000023-NQSTI project by the Italian Ministry of University and Research, co-funded by the European Union - NextGeneration EU, and from Provincia Autonoma di Trento (PAT), partly via the Q@TN initiative. I.C.'s research was supported in part by the International Centre for Theoretical Sciences (ICTS) for participating in the program: Turbulence and Vortex dynamics in 2D quantum fluids (code: ICTS/QUFLU2024/2). A.P. thanks fruitful discussions with P. B\'ejot and B. Kibler.

\newpage
\centerline{{\bf APPENDIX}}

\section{Derivation of the general form of UPE (2)}

In this section we derive the general form of the unidirectional propagation equation (2)  for a multimode waveguide.

We start from the Maxwell equations and consider the scalar approximation for the real-valued electric field $E(t,z,\br)$.
Considering the cubic (Kerr) nonlinearity, the Fourier transform of the field $\hat{E}({\tilde \omega},z,\br) = \int E(t,z,\br)e^{i{\tilde \omega} t} dt$
satisfies the nonlinear Helmholtz equation:
\begin{equation}
\label{eq:helm}
\Delta \hat E + \frac{{\tilde \omega}^2}{c^2}  {n}^2({\tilde \omega},\br) \hat E  = -\frac{{\tilde \omega}^2}{c^2} \chi^{(3)} \widehat{E^3} ,
\end{equation}
where $\Delta =\Delta_\perp+\partial_z^2$ is the three-dimensional Laplacian and $c$ is the vacuum light speed.
Here, $n({\tilde \omega},\br)$ is the index of refraction of the waveguide at the frequency ${\tilde \omega}$ (it does not depend on $z$ which is the waveguide axis). We denote $n({\tilde \omega})={n}({\tilde \omega},{\bf 0})$ the index of refraction at $\br={\bf 0}$, and $n_o=n({\tilde \omega}_o)$, with ${\tilde \omega}_o$ the central frequency of the field.
We may have ${n}^2({\tilde \omega},\br) = {n}^2({\tilde \omega})( 1-|\br|^2/a^2)$ for a parabolic (graded-index) fiber, or ${\tilde n}^2({\tilde \omega},\br) ={n}^2({\tilde \omega})$ 
 for a step-index (finite-size homogeneous) waveguide.
The wavenumber at frequency ${\tilde \omega}$ is $k({\tilde \omega})=n({\tilde \omega}) {\tilde \omega}/c$.

We introduce the real-valued orthonormal basis $u_m(\br)$, with eigenvalues $\beta_m$, which are solutions of
$$
-\frac{1}{2k_o} \Delta_\perp u_m + V(\br)
  u_m=\beta_m u_m,
$$
where $k_o=k({\tilde \omega}_o)= n_o{\tilde \omega}_o/c$ and $V(\br)= \frac{{\tilde \omega}_o}{2n_oc}\big(n_o^2-n({\tilde \omega}_o,\br)^2  \big)$.
The field  can be expanded on this basis
$\hat E({\tilde \omega},z,\br) =\sum_m {a}_m({\tilde \omega},z) u_m(\br)$,
where ${a}_m({\tilde \omega},z) = \int u_m(\br) \hat E({\tilde \omega},z,\br) d\br$ (remember that the $u_m$ are real-valued).
Note that ${a}_m(-{\tilde \omega},z)={a}_m^*({\tilde \omega},z)$  (where the star means complex conjugate) because $E(t,z,\br)$ is real-valued.

By multiplying the Helmholtz Eq.(\ref{eq:helm}) by $u_m$ and integrating in $\br$, 
we get 
\begin{align}
&  \partial_z^2 {a}_m +\big(  n_o^2{\tilde \omega}^2/c^2   -  2k_o \beta_m\big)   {a}_m + \sum_p A_{mp}({\tilde \omega}) {a}_p  \nonumber \\
&  =-\frac{\chi^{(3)} {\tilde \omega}^2}{(2\pi)^2 c^2}   \sum_{p,q,s} W_{mpqs} \int {a}_p({\tilde \omega}_1){a}_q({\tilde \omega}_2){a}_s({\tilde \omega}_3) {\hat \delta}_{\tilde \omega} d{\tilde \omega}_{123},
\label{eq:helm_am_gen}
\end{align}
with $W_{mpqs} = \int u_m(\br)u_p(\br)u_q(\br)u_s(\br) d\br$, $\hat{\delta}_{\tilde \omega}=\delta({\tilde \omega}-{\tilde \omega}_1-{\tilde \omega}_2-{\tilde \omega}_3)$, $d{\tilde \omega}_{123}=d{\tilde \omega}_1 d{\tilde \omega}_2 d{\tilde \omega}_3$, and 
\begin{align*}
A_{mp}({\tilde \omega})&=  \int Q({\tilde \omega},\br) u_m(\br) u_p(\br) d\br   ,\\
Q({\tilde \omega},\br)&= \frac{{\tilde \omega}^2}{c^2}\big({n}^2({\tilde \omega},\br) -n_o^2\big) - \frac{{\tilde \omega}_o^2}{c^2}  \big({n}^2({\tilde \omega}_o,\br) -n_o^2\big).
\end{align*}
In general, the coefficients ${a}_m$ have rapid dominant phases of the form $\exp(i k_o z -i \beta_m z)$, so the terms $p\neq m$ in the sum $\sum_p A_{mp}({\tilde \omega}) {a}_p$ average out.
The resulting equation can be written
\begin{align}
&  \partial_z^2 {a}_m +\big(  n_o^2{\tilde \omega}^2/c^2    -  2k_o \beta_m+\alpha_m ({\tilde \omega})\big)     {a}_m 
\nonumber \\
&= -\frac{\chi^{(3)} {\tilde \omega}^2 }{(2\pi)^2 c^2}   \sum_{p,q,s} W_{mpqs} \int {a}_p({\tilde \omega}_1) {a}_q^*({\tilde \omega}_2) {a}_s({\tilde \omega}_3) {\delta}_{\tilde \omega} d{\tilde \omega}_{123},
\label{eq:helm_am}
\end{align}
with $\alpha_m({\tilde \omega}) = A_{mm}({\tilde \omega})$. Note that we have changed ${\tilde \omega}_2 \to -{\tilde \omega}_2$, so that the Dirac function reads
$\delta_{\tilde \omega}=\delta({\tilde \omega}-{\tilde \omega}_1+{\tilde \omega}_2-{\tilde \omega}_3)$.
Notice that, at variance with the multimode UPE in Ref.\cite{bejot19}, here $\beta_m$ and $W_{mpqr}$ do not depend on the frequency ${\tilde \omega}$, because we have expanded the field on the eigenmodes at frequency ${\tilde \omega}_o$. Then we just need to compute $\beta_m$, $\alpha_m({\tilde \omega})$ and $W_{mpqs}$, and not a function indexed by $m,p,q,s$ and four frequencies. The latter would have been necessary if we had chosen to expand the field on ${\tilde \omega}$-dependent modes.

\bigskip \noindent
{\bf Application to a homogeneous trapping potential:}  
In the case of a step-index waveguide featured by a  homogeneous trapping potential (e.g., a circular waveguide of radius $R$ with $V(\br)=0$ for $|\br| \le R$ and $V(\br)=V_0$ for $|\br| > R$), Eq.(\ref{eq:helm_am}) is obtained without approximations from Eq.(\ref{eq:helm_am_gen}). Indeed, for a step-index waveguide the matrix $A_{mp}({\tilde \omega})$ is diagonal, with 
\begin{align}
&A_{mp}({\tilde \omega})=\alpha({\tilde \omega}) \delta_{mp}, \\ 
&\alpha({\tilde \omega})=({\tilde \omega}^2/c^2) \big( n({\tilde \omega})^2-  n_o^2 \big).
\end{align}

\bigskip \noindent
{\bf The forward scattering approximation:}
Eq.(\ref{eq:helm_am}) can be written as 
\begin{align}
&  \partial_z^2 {a}_m +\big(  k_o-{\hat \beta}_m({\tilde \omega}) \big)^2    {a}_m = -F_m[{\bm a}] ,
\label{eq:helm_am2}
\end{align}
with 
\begin{equation}
\label{eq:defBn2}
{\hat \beta}_m({\tilde \omega})=k_o - \sqrt{  n_o^2{\tilde \omega}^2/c^2       -  2k_o \beta_m    + \alpha_m({\tilde \omega})  }  
\end{equation}
and
\begin{equation}
F_m[{\bm a}]= \frac{\chi^{(3)} {\tilde \omega}^2 }{(2\pi)^2 c^2}   \sum_{p,q,s} W_{mpqs} \int {a}_p({\tilde \omega}_1) {a}_q^*({\tilde \omega}_2) {a}_s({\tilde \omega}_3) \delta_{\tilde \omega} d{\tilde \omega}_{123}.
\end{equation}
Note that ${\hat \beta}_m({\tilde \omega}_o) \neq \beta_m$ in general.

We first remark that, when there is no nonlinearity $\chi^{(3)}=0$,  the general solution of Eq.(\ref{eq:helm_am2})  has the form 
$$
{a}_m({\tilde \omega},z)= {a}_m^+({\tilde \omega}) e^{ i (k_o  -{\hat \beta}_m({\tilde \omega})) z} +{a}_m^-({\tilde \omega}) e^{- i (k_o  -{\hat \beta}_m({\tilde \omega})) z}  ,
$$
where ${a}_m^+({\tilde \omega}), {a}_m^-({\tilde \omega})$ do not depend on $z$ and are determined by the source or initial conditions.

In the presence of the nonlinearity we introduce
\begin{align*}
{a}_m^+({\tilde \omega},z) &= \frac{1}{2}\Big(  {a}_m({\tilde \omega},z)   + \frac{\partial_z {a}_m({\tilde \omega},z) }{i(k_o-{\hat \beta}_m({\tilde \omega}))} \Big) e^{-i(k_o-{\hat \beta}_m({\tilde \omega}))z} 
, \\
{a}_m^-({\tilde \omega},z) &= \frac{1}{2}\Big(  {a}_m({\tilde \omega},z)   - \frac{\partial_z {a}_m({\tilde \omega},z)}{i(k_o-{\hat \beta}_m({\tilde \omega}))}  \Big) e^{i(k_o-{\hat \beta}_m({\tilde \omega}))z} 
,
\end{align*}
so that the solution of Eq.(\ref{eq:helm_am2}) has the form
\begin{align*}
{a}_m({\tilde \omega},z) = {a}_m^+({\tilde \omega},z) e^{ i (k_o  -{\hat \beta}_m({\tilde \omega})) z } +{a}_m^-({\tilde \omega},z) e^{- i (k_o  -{\hat \beta}_m({\tilde \omega})) z }  ,
\end{align*}
and ${a}_m^+({\tilde \omega},z) $ and ${a}_m^-({\tilde \omega},z) $
satisfy the coupled first-order system
\begin{align}
\partial_z {a}_m^+ &=  \frac{i}{2(k_o-{\hat \beta}_m({\tilde \omega}))} F_m [{\bm a}] e^{-i(k_o-{\hat \beta}_m({\tilde \omega}))z} ,\\
\partial_z {a}_m^- &= -\frac{i}{2(k_o-{\hat \beta}_m({\tilde \omega}))} F_m [{\bm a}] e^{+i(k_o-{\hat \beta}_m({\tilde \omega}))z} .
\end{align}
Neglecting the backscattered components ${a}_m^-$, introducing the mode amplitudes defined by 
\begin{equation}
{b}_m(\omega,z) = {a}_m({\tilde \omega}_o+{\omega},z) e^{-i k_oz - i \frac{{\omega}}{v_g} z} ,
\end{equation} 
(which means we observe the field in a reference frame that propagates at the group velocity $v_g^{-1}=\partial_{\tilde \omega} k({\tilde \omega}_o)$),
we obtain the general form of the UPE:
\begin{align}
&i \partial_z {b}_m({\omega},z) = {\tilde \beta}_m({\omega})   {b}_m  -  \gamma \Gamma_m({\omega})   P_{m}[{\bm b}],
\label{eq:upe1}\\
&P_{m}[{\bm b}]=\sum_{pqs} W_{mpqs} \int {b}_p({\omega}_1) {b}_q^*({\omega}_2){b}_s({\omega}_3) \delta_{{\omega}} 
d{\omega}_{123},
\label{eq:upe2}
\end{align}
with ${\delta}_{{\omega}}=\delta({\omega}-{\omega}_1+{\omega}_2-{\omega}_3)$, 
\begin{align}
{\tilde \beta}_m({\omega})&={\hat \beta}_m({\tilde \omega}_o+{\omega})+\frac{{\omega}}{v_g},\\
\Gamma_m({\omega}) &= \frac{({\tilde \omega}_o+{\omega})^2}{{\tilde \omega}_o^2} \frac{ k_o}{k_o-{\tilde \beta}_m({\tilde \omega}_o+{\omega})},
\end{align}
and $\gamma = \chi^{(3)} {\tilde \omega}_o/(8\pi^2 c)$.

The UPE (\ref{eq:upe1}-\ref{eq:upe2}) conserves the Hamiltonian $H_{b}=E_{b}+U_{b}$:
\begin{align}
E_{b} &=\sum_m \int \frac{{\tilde \beta}_m({\omega})}{\Gamma_m({\omega})} |{b}_m({\omega})|^2 d{\omega}, 
\label{eq:H_E_b}\\
U_{b} &= -\frac{\gamma}{2} \sum_{mpqs}  W_{mpqs}\int
{b}_m^*({\omega}_1) {b}_p({\omega}_2) {b}_q^*({\omega}_3) {b}_s({\omega}_4) \delta_{1234}
d{\omega}_{1234}
\label{eq:H_U_b}
\end{align}
with  $\delta_{1234}=\delta({\omega}_1+{\omega}_3-{\omega}_2-{\omega}_4)$.
It also conserves the particle number and the `temporal' momentum 
\begin{align}
N_{b} = \sum_m  \int \frac{|{b}_m({\omega})|^2}{\Gamma_m({\omega})} d{\omega},  \ 
P_{b} = \sum_m   \int {\omega} \frac{|{b}_m({\omega})|^2}{\Gamma_m({\omega})} d{\omega}.
\label{eq:N_b_P_b}
\end{align}

\bigskip
\noindent
{\bf The NLSE approximation:}
Making use of the slowly-varying envelope approximation in the time domain and the paraxial approximation in the space domain, and using 
$k({\tilde \omega}_o+{\omega})=k({\tilde \omega}_o) + \partial_{\tilde \omega} k({\tilde \omega}_o) {\omega}+\frac{1}{2}\partial_{\tilde \omega}^2 k({\tilde \omega}_o){\omega}^2 $, gives
${\tilde \beta}_m( {\omega}) = \beta_m -\kappa_2 {\omega}^2$,  
because $\frac{1}{v_g}=\partial_{\tilde \omega} k({\tilde \omega}_o)$ and $\kappa_2=\frac{1}{2}\partial_{\tilde \omega}^2 k({\tilde \omega}_o)$.
The nonlinear dispersion relation reduces to a constant, $\Gamma_m({\omega})=1$.
The slowly varying envelope of the field $E(t,\br,z)$ observed in the moving frame
\begin{align}
\nonumber
\psi(t,\br,z) &= E \Big(t +\frac{z}{v_g},\br,z\Big)e^{-ik_o z+i{\tilde \omega}_o(t +\frac{z}{v_g})} \\
\nonumber
&= \frac{1}{2\pi}  \int \hat{E}({\tilde \omega}_o+{\omega} ,\br,z) e^{-ik_o z -i {\omega} (t + \frac{z}{v_g}) } d{\omega} \\
&= \frac{1}{2\pi} \sum_m \int b_m({\omega} ,z) e^{-i {\omega} t } d{\omega} u_m(\br) 
\label{eq:defpsi:app}
\end{align}
then satisfies the NLSE (1), with $\gamma_o=(2\pi)^2\gamma$.
The nonlinear length is defined by $L_{\rm nl}=1/(|\gamma_o| {\bar N}_{b})$, where ${\bar N}_{b}=N_b/({\cal T} A_{\rm eff})$ is the optical intensity with $N_b/{\cal T}$ the optical power (averaged over the numerical time window ${\cal T}$), and $A_{\rm eff} = 1/W_{0000}$ the effective area of the fundamental mode.

\bigskip
\noindent
{\bf The NEE approximation:}
The main difference with respect to the NLSE limit is that the NEE preserves the frequency dependence of $k({\tilde \omega})$, but neglects the dispersion of the refractive index, i.e., we set ${n}({\tilde \omega})={n}({\tilde \omega}_o)$ in $k({\tilde \omega})={n}({\tilde \omega}_o){\tilde \omega}/c$ \cite{kolesik04}, so that  $\frac{2k_o \beta_m}{2 k({\tilde \omega}_o+{\omega})}\simeq \frac{\beta_m}{1+\frac{{\omega}}{{\tilde \omega}_o}}$, and
\begin{equation}
{\tilde \beta}_m({\omega}) = \frac{\beta_m}{1+\frac{{\omega}}{{\tilde \omega}_o}} - \sum_{j\ge 2} \kappa_j {\omega}^j ,
\label{eq:limit_step_nee}
\end{equation}
with $\kappa_j=\frac{1}{j!}\partial_{\tilde \omega}^j k({\tilde \omega}_o)$.
The first term in Eq.(\ref{eq:limit_step_nee}) couples spatial and temporal effects.
The approximation ${\tilde \beta}_m(\omega)\simeq \beta_m\big( 1-\frac{{\omega}}{{\tilde \omega}_o}\big) - \sum_{j\ge 2} \kappa_j {\omega}^j$  is called partially-corrected NLS equation \cite{kolesik04}.\\
For the nonlinear coefficient $\Gamma_m({\omega})$, the NEE approximates $\sqrt{  k({\tilde \omega}_o+{\omega})^2  -  2k_o \beta_m } \simeq k({\tilde \omega}_o+{\omega})$. 
Neglecting the dispersion of the refractive index, we get 
\begin{equation}
\Gamma_m({\omega}) \simeq \frac{({\tilde \omega}_o+{\omega})^2}{{\tilde \omega}_o^2} \frac{k_o}{{n}({\tilde \omega}_o)({\tilde \omega}_o+{\omega})/c}=
 1+ \frac{{\omega}}{{\tilde \omega}_o}  .
\label{eq:Gamm_m_nee}
\end{equation}

Note that the NEE can be written in the space-time domain.
Considering the complex envelope field $\psi(t,\br,z)$ defined by Eq.(\ref{eq:defpsi:app}), the NEE reads
\begin{align}
i\partial_z \psi(t,\br,z) &= \frac{-1}{2k_o(1+i{\tilde \omega}_o^{-1} \partial_t)}
\nabla^2 \psi 
+\frac{1}{(1+i{\tilde \omega}_o^{-1} \partial_t)} V(\br) \psi \nonumber \\
& \quad +\sum_{j\ge 2} \kappa_j \partial_t^j \psi 
- \gamma_o \big(1+i{\tilde \omega}_o^{-1} \partial_t \big) |\psi|^2\psi.
\label{eq:nee_rt}
\end{align}
By removing the operators $(1+i{\tilde \omega}_o^{-1} \partial_t)$ and by retaining only second-order dispersion effects ($j=2$), Eq.(\ref{eq:nee_rt}) recovers the NLSE approximation.

\widetext
\section{Derivation of the wave turbulence kinetic equation}
\label{sec:deriv_KE}

In this section we derive the wave turbulence kinetic equation from the general form of the UPE (\ref{eq:upe1}-\ref{eq:upe2}). The derived kinetic equation then includes, as particular case, the kinetic Eq.(3) in the NLSE approximation.

We write the UPE (\ref{eq:upe1}-\ref{eq:upe2}) in a symmetric form by introducing the mode amplitudes
$\check{b}_m(\omega,z) = {b}_m(\omega,z) / \sqrt{\Gamma_m(\omega)}$, which satisfy
\begin{equation}
i \partial_z \check{b}_m = {\tilde \beta}_m(\omega)   \check{b}_m  -   \gamma \sum_{p,q,s} \iiint  L_{mpqs} (\omega,\omega_1,\omega_2,\omega_3)\check{b}_p(\omega_1) {\check{b}_q^*(\omega_2)}\check{b}_s(\omega_3) \delta(\omega-\omega_1+\omega_2-\omega_3) d\omega_1 d\omega_2 d\omega_3 ,
\label{eq:evol4}
\end{equation}
with  
$$
L_{mpqs}(\omega,\omega_1,\omega_2,\omega_3)
=
W_{mpqs} \sqrt{ \Gamma_m(\omega)\Gamma_p(\omega_1)\Gamma_q(\omega_2)\Gamma_s(\omega_3)}.
$$
Equation (\ref{eq:evol4}) conserves the Hamiltonian $H_{\check b}=E_{\check b}+U_{\check b}$,
with:
$E_{\check b} =\sum_m \int {\tilde \beta}_m(\omega) |\check b_m(\omega)|^2 d\omega$, and 
\begin{align}
U_{\check b} = -\frac{\gamma}{2} \sum_{mpqs} \iiiint L_{mpqs}(\omega,\omega_1,\omega_2,\omega_3)
{\check b}_m^*(\omega) {\check b}_p(\omega_1) {\check b}_q^*(\omega_2) {\check b}_s(\omega_3) \delta(\omega-\omega_1+\omega_2-\omega_3)
d\omega d\omega_1 d\omega_2 d\omega_3,
\label{eq:U_st}
\end{align}
as well as the particle number 
$N_{\check b} = \sum_m  \int |{\check b}_m(\omega)|^2 d\omega$, and `temporal' momentum 
$P_{\check b} = \sum_m   \int \omega |{\check b}_m(\omega)|^2 d\omega$.

We consider random initial conditions with statistically stationary (in time) distribution.
The initial spectrum $n_m(\omega,0)$ is such that $\left< {b}_m(\omega, 0) {b}_{p}^*(\omega', 0) \right> ={n}_m(\omega, 0) \delta_{mp} \delta(\omega-\omega')$.
The brackets $\left< \cdot \right>$ denote an average with respect to the distribution of the random initial conditions at $z=0$.
We have
\begin{align*}
&\partial_z \left< \check{b}_m(\omega_1) {\check{b}_m^*(\omega_1')}\right>
= i \big({\tilde \beta}_m(\omega_1') -{\tilde \beta}_m(\omega_1) \big)  \left< \check{b}_m(\omega_1) {\check{b}_m^*(\omega_1')}\right>\\
&+
{i \gamma }  \iiint \sum_{p,q,s} L_{mpqs}(\omega_1',\omega_2,\omega_3,\omega_4)
 \left<{\check{b}_m^*(\omega_1')}  \check{b}_{p}(\omega_2)
 { \check{b}_{q}^*(\omega_3)}  \check{b}_{s}(\omega_4)  \right>
\delta(\omega_2-\omega_3+\omega_4-\omega_1) d\omega_{2,3,4} \\
&-
 {i \gamma}   \iiint \sum_{p,q,s} L^*_{mpqs}(\omega_1,\omega_2,\omega_3,\omega_4)
 \left<{\check{b}_m(\omega_1)}  \check{b}_{p}^*(\omega_2)
{ \check{b}_{q}(\omega_3)}  {\check{b}_{s}^*(\omega_4) } \right>
\delta(\omega_2-\omega_3+\omega_4-\omega_1') d\omega_{2,3,4}.
\end{align*}
and
\begin{align*}
&\partial_z \left< {\check{b}_m^*(\omega_1)}
\check{b}_{p}(\omega_2) {\check{b}_{q}^*(\omega_3)} \check{b}_{s}(\omega_4) \right>
= i \Omega_{mpqs}(\omega_1,\omega_2,\omega_3,\omega_4) \left< {\check{b}_m^*(\omega_1)}
\check{b}_{p}(\omega_2) {\check{b}_{q}^*(\omega_3)} \check{b}_{s}(\omega_4) \right>
\\
&
-  {i\gamma }  \sum_{p',q',s'}
\iiint  
L_{mp'q's'}^*(\omega_1,\omega_1',\omega_2',\omega_3')
\left< {\check{b}_{p'}^*(\omega_1')}
\check{b}_{q'}(\omega_2') {\check{b}_{s'}^*(\omega_3')} \check{b}_{p}(\omega_2)  {\check{b}_q^*(\omega_3)}
\check{b}_{s}(\omega_4) \right>
\delta(\omega_1'-\omega_2'+\omega_3'-\omega_1) d\omega_{1,2,3}' 
\\
&
+  {i \gamma}   \sum_{p',q',s'}
\iiint 
{L_{pp'q's'}}(\omega_2,\omega_1',\omega_2',\omega_3')
 \left< {\check{b}_m^*(\omega_1)}
\check{b}_{p'}(\omega_1') {\check{b}_{q'}^*(\omega_2')} \check{b}_{s'}(\omega_3')  {\check{b}_q^*(\omega_3)}
\check{b}_{s}(\omega_4) \right>
\delta(\omega_1'-\omega_2'+\omega_3'-\omega_2) d\omega_{1,2,3}' \\
&
- {i \gamma}   \sum_{p',q',s'}
 \iiint 
L_{qp'q's'}^*(\omega_3,\omega_1',\omega_2',\omega_3')
 \left< {\check{b}_m^*(\omega_1)}
\check{b}_p(\omega_2) {\check{b}_{p'}^*(\omega_1')} \check{b}_{q'}(\omega_2')  {\check{b}_{s'}^*(\omega_3')}
\check{b}_{s}(\omega_4) \right>
\delta(\omega_1'-\omega_2'+\omega_3'-\omega_3) d\omega_{1,2,3}'\\
&
+  {i \gamma}  \sum_{p',q',s'}
 \iiint 
{L_{sp'q's'}}(\omega_4,\omega_1',\omega_2',\omega_3')
 \left< {\check{b}_m^*(\omega_1)}
\check{b}_p(\omega_2) {\check{b}_q^*(\omega_3)} \check{b}_{p'}(\omega_1')  {\check{b}_{q'}^*(\omega_2')}
\check{b}_{s'}(\omega_3') \right>
\delta(\omega_1'-\omega_2'+\omega_3'-\omega_4) d\omega_{1,2,3}',
\end{align*}
with
$\Omega_{mpqs}(\omega_1,\omega_2,\omega_3,\omega_4)  =
 {\tilde \beta}_m(\omega_1)-{\tilde \beta}_p(\omega_2)+{\tilde \beta}_q(\omega_3)-{\tilde \beta}_s(\omega_4)$. 
 The equation for the evolution of the fourth-order moment (in the rhs of the above equation) will depend on the sixth-order moment. In this way, one obtains an infinite hierarchy of moment equations, in which the $n-$th order moment depends on the $n + 2$-order moment. 
The hierarchy is closed in the weakly nonlinear regime because the field approaches Gaussian statistics.
By virtue of the factorizability property of statistical Gaussian fields, we obtain 
\begin{align*}
&\partial_z \left< {\check{b}_m^*(\omega_1)}
\check{b}_{p}(\omega_2) {\check{b}_{q}^*(\omega_3)} \check{b}_{s}(\omega_4) \right>
= i \Omega_{mpqs}(\omega_1,\omega_2,\omega_3,\omega_4) \left< {\check{b}_m^*(\omega_1)}
\check{b}_{p}(\omega_2) {\check{b}_{q}^*(\omega_3)} \check{b}_{s}(\omega_4) \right>
\\
&
- {2 i\gamma }  
L^*_{mpqs}(\omega_1,\omega_2,\omega_3,\omega_4) {\check n}_m(\omega_1) {\check n}_p(\omega_2){\check n}_q(\omega_3){\check n}_s(\omega_4) \\
&\qquad 
\times \big[ {\check n}_m(\omega_1)^{-1}-{\check n}_p(\omega_2)^{-1}
+{\check n}_q(\omega_3)^{-1}-{\check n}_s(\omega_4)^{-1}\big] \delta(\omega_3-\omega_4+\omega_1-\omega_2)
\\
&
- {2 i\gamma } \Big[ \sum_{p'}
 \int L^*_{mpp'p'} (\omega_1,\omega_1,\omega,\omega) {\check n}_{p'}(\omega)d\omega\Big] \big( {\check n}_p(\omega_1)-{\check n}_m(\omega_1)\big) {\check n}_q(\omega_3) \delta_{qs} \delta(\omega_2-\omega_1) \delta(\omega_3-\omega_4+\omega_1-\omega_2)\\
&
- {2 i\gamma } \Big[ \sum_{p'}
\int L^*_{mp'p's}(\omega_1,\omega,\omega,\omega_1)  {\check n}_{p'}(\omega)d\omega\Big] \big( {\check n}_s(\omega_1)-{\check n}_m(\omega_1)\big) {\check n}_p(\omega_2) \delta_{pq} \delta(\omega_4-\omega_1) \delta(\omega_3-\omega_4+\omega_1-\omega_2)
\\
&
- {2 i\gamma } \Big[ \sum_{p'}
 \int L^*_{p'p'qs}(\omega,\omega,\omega_3,\omega_3) {\check n}_{p'}(\omega)d\omega\Big] \big( {\check n}_s(\omega_3)-{\check n}_q(\omega_3)\big) {\check n}_m(\omega_1) \delta_{pm} \delta(\omega_4-\omega_3) \delta(\omega_3-\omega_4+\omega_1-\omega_2)
\\
&
- {2 i\gamma } \Big[ \sum_{p'}
 \int L^*_{p'p'qp}(\omega,\omega,\omega_2,\omega_2) {\check n}_{p'}(\omega)d\omega\Big] \big( {\check n}_p(\omega_2)-{\check n}_q(\omega_2)\big) {\check n}_m(\omega_1) \delta_{sm} \delta(\omega_3-\omega_2) \delta(\omega_3-\omega_4+\omega_1-\omega_2)
 .
\end{align*}
where $\left< \check{b}_m(\omega_1) {\check{b}_{m'}^*}(\omega_1') \right> ={\check n}_m(\omega_1) \delta_{mm'} \delta(\omega_1-\omega_1')$.
By the presence of the factor $ \delta(\omega_3-\omega_4+\omega_1-\omega_2)$ in the rhs, we have
$$
 \left< {\check{b}_m^*(\omega_1)}
\check{b}_{p}(\omega_2) {\check{b}_{q}^*(\omega_3)} \check{b}_{s}(\omega_4) \right>
= \delta(\omega_3-\omega_4+\omega_1-\omega_2)
J_{mpqs}(\omega_1,\omega_2,\omega_3,\omega_4)
$$
and the spectrum ${\check n}_m(\omega_1)$ satisfies
\begin{align}
& \partial_z {\check n}_m(\omega_1) = -{2\gamma}  \sum_{p,q,s} {\rm Im}\Big(   \iiint L_{mpqs}(\omega_1,\omega_2,\omega_3,\omega_4) {J}_{mpqs}(\omega_1,\omega_2,\omega_3,\omega_4)\delta(\omega_1+\omega_3-\omega_2-\omega_4) d\omega_{2,3,4}  \Big) ,
\label{eq:n_m_1}
\\
\nonumber
& \partial_z {J}_{mpqs}(\omega_1,\omega_2,\omega_3,\omega_4) = i \Omega_{mpqs}(\omega_1,\omega_2,\omega_3,\omega_4) {J}_{mpqs}(\omega_1,\omega_2,\omega_3,\omega_4) \\
 &\qquad - {2 i\gamma } L^*_{mpqs}(\omega_1,\omega_2,\omega_3,\omega_4)  M_{mpqs}[{\bm {\check n}}(z)](\omega_1,\omega_2,\omega_3,\omega_4)  - {2 i \gamma}  R_{mpqs}[{\bm {\check n}}(z)](\omega_1,\omega_2,\omega_3,\omega_4)  ,
\label{eq:J_m_1}
 \end{align}
 where
\begin{align}
M_{mpqs}  [{\bm {\check n}}](\omega_1,\omega_2,\omega_3,\omega_4) =& 
{\check n}_m(\omega_1) {\check n}_p(\omega_2) {\check n}_q(\omega_3) {\check n}_s(\omega_4) ({\check n}_m^{-1}(\omega_1)+{\check n}_q^{-1}(\omega_3)-{\check n}_p^{-1}(\omega_2)-{\check n}_s(\omega_4)^{-1})  ,
\\
\nonumber
R_{mpqs}  [{\bm {\check n}}](\omega_1,\omega_2,\omega_3,\omega_4) =& 
{{\cal U}_{mp}^*[{\bm {\check n}}]}(\omega_1) \big( {\check n}_p(\omega_2)-{\check n}_m(\omega_1) \big) {\check n}_q(\omega_3) \delta_{qs} \delta(\omega_4-\omega_3) \\
\nonumber
&+{{\cal U}_{ms}^*[{\bm {\check n}}]}(\omega_1) \big( {\check n}_s(\omega_4)-{\check n}_m(\omega_1) \big) {\check n}_p(\omega_2) \delta_{pq} \delta(\omega_3-\omega_2) \\
\nonumber
&+{{\cal U}_{qs}^*[{\bm {\check n}}]}(\omega_3) \big( {\check n}_s(\omega_4)-{\check n}_q(\omega_3) \big) {\check n}_m(\omega_1) \delta_{pm} \delta(\omega_2-\omega_1) 
\\
&+{{\cal U}_{qp}^*[{\bm {\check n}}]}(\omega_3) \big( {\check n}_p(\omega_2)-{\check n}_q(\omega_3) \big) {\check n}_m(\omega_1) \delta_{sm} \delta(\omega_4-\omega_1) ,
\nonumber\\
{\cal U}_{qp}[{\bm {\check n}}](\omega) =& \sum_{p'} \int L_{qpp'p'} (\omega,\omega,\omega',\omega') {\check n}_{p'}(\omega')d\omega' .
\end{align}
After simplification,
we finally obtain
\begin{align}
 \partial_z {\check n}_m(\omega_1) =&  -{2\gamma}    {\rm Im}\Big( \sum_{p,q,s}  \iint 
 {J}_{mpqs}^{(1)}(\omega_1,\omega_2,\omega_3) d\omega_{2,3}  
 + 2\sum_p J^{(2)}_{mp}(\omega_1) \Big) ,
\label{eq:kin2a}
 \end{align}
 with
 \begin{align}
\nonumber
\partial_z {J}_{mpqs}^{(1)}(\omega_1,\omega_2,\omega_3) =& i \Omega_{mpqs}(\omega_1,\omega_2,\omega_3,\omega_1+\omega_3-\omega_2) {J}_{mpqs}^{(1)}(\omega_1,\omega_2,\omega_3) \\
 &- {2 i \gamma} |{L_{mpqs}}(\omega_1,\omega_2,\omega_3,\omega_1+\omega_3-\omega_2)|^2 M_{mpqs}[{\bm {\check n}}(z)](\omega_1,\omega_2,\omega_3,\omega_1+\omega_3-\omega_2)   ,
 \label{eq:kin2b}
 \end{align}
and
\begin{align}
 \partial_z {J}_{mp}^{(2)}(\omega_1) =& i\big({\tilde \beta}_m(\omega_1)-{\tilde \beta}_p(\omega_1) \big) {J}_{mp}^{(2)}(\omega_1) 
 - {2 i \gamma} |{\cal U}_{mp}[{\bm {\check n}}(z)] (\omega_1)|^2 \big( {\check n}_p(\omega_1)-{\check n}_m(\omega_1)\big).
 \label{eq:kin2c}
\end{align}
Eqs.(\ref{eq:kin2a}-\ref{eq:kin2c}) are the equations driving the evolution of the spectrum.

Making use of the wave turbulence theory \cite{Newell-Rumpf,nazarenko11}, and neglecting degenerate modes, we obtain the kinetic equation governing the evolution of the modal spectra ${\check n}_m(\omega,z)$:
\begin{align}
\nonumber
&\partial_z {\check n}_m(\omega_1) =  {4\pi \gamma^2}   \sum_{p,q,s}    \iiint |L_{mpqs}(\omega_1,\omega_2,\omega_3,\omega_4)|^2
 M_{mpqs}[{\bm {\check n}}](\omega_1,\omega_2,\omega_3,\omega_4) \\
 &\qquad \qquad \qquad  \times 
\delta(\omega_1+\omega_3-\omega_2-\omega_4)  \delta\big( \Omega_{mpqs}(\omega_1,\omega_2,\omega_3,\omega_4) \big) d\omega_2 d\omega_3 d\omega_4   
\label{eq:kin_eqn}\\
&M_{mpqs}  [{\bm {\check n}}](\omega_1,\omega_2,\omega_3,\omega_4) =
{\check n}_m(\omega_1) {\check n}_p(\omega_2) {\check n}_q(\omega_3) {\check n}_s(\omega_4) ({\check n}_m^{-1}(\omega_1)+{\check n}_q^{-1}(\omega_3)-{\check n}_p^{-1}(\omega_2)-{\check n}_s(\omega_4)^{-1})  ,\\
&\Omega_{mpqs}(\omega_1,\omega_2,\omega_3,\omega_4)  =
 {\tilde \beta}_m(\omega_1)-{\tilde \beta}_p(\omega_2)+{\tilde \beta}_q(\omega_3)-{\tilde \beta}_s(\omega_4)\\
&L_{mpqs}(\omega_1,\omega_2,\omega_3,\omega_4)=W_{mpqs} \sqrt{ \Gamma_m(\omega_1)\Gamma_p(\omega_2)\Gamma_q(\omega_3)\Gamma_s(\omega_4)}
\label{eq:Omega_mpqs}
 \end{align}
The kinetic Eq.(\ref{eq:kin_eqn}) conserves the 
particle number $N_{\check{n}} = \frac{1}{(2\pi)^2}\sum_m  \int \check{n}_m(\omega) d\omega$, the momentum $P_{\check{n}} = \frac{1}{(2\pi)^2} \sum_m   \int \omega \check{n}_m(\omega) d\omega$, and the kinetic energy  $E_{\check{n}} =\frac{1}{(2\pi)^2}\sum_m \int {\tilde \beta}_m(\omega) \check{n}_m(\omega) d\omega$. 
Note that these quantities are related to the definitions of $N_{\check b}$, $P_{\check b}$, $E_{\check b}$ given above around Eq.(\ref{eq:U_st}) but they have been renormalized to take into account that the field is here statistically stationary in time so that the quantities $N_{\check b}$, $P_{\check b}$, $E_{\check b}$  are infinite while the quantities $N_{\check{n}}$, $E_{\check{n}}$, $E_{\check{n}}$ are well-defined and finite.\\
Eq.(\ref{eq:kin_eqn}) exhibits an $H-$theorem of entropy growth $\partial_z S_{\check n}(z) \ge 0$,  for the nonequilibrium entropy $S_{\check n}(z)=\sum_m  \int \log\big(\check{n}_m(\omega)\big) d\omega$.
Then Eq.(\ref{eq:kin_eqn}) describes the nonequilibrium process of ST thermalization toward equilibrium.
In terms of the original variables $n_m(\omega)$, the equilibrium ST distribution has a Rayleigh-Jeans (RJ) form 
\begin{equation}
n_m^{RJ}( {{\omega}})
=\check{n}_m^{RJ}( {{\omega}}) \Gamma_m( {{\omega}})
= \frac{T \, \Gamma_m( {{\omega}})}{{\tilde \beta}_m( {{\omega}}){-\lambda {\omega}}-\mu} ,
\label{eq:n_rj_general}
\end{equation}
where the temperature $T$, chemical potential $\mu$, and average `velocity' $\lambda$, are determined from the three conserved quantities 
($N_{\check{n}}, P_{\check{n}}, E_{\check{n}}$).

\medskip
\noindent
{\bf NLSE approximation:} In the NLSE case, we set $\Gamma_m(\omega)=1, {\tilde \beta}_m(\omega)=\beta_m - \kappa_2 \omega^2$, so that $L_{mpqr}^{\omega_1 \omega_2 \omega_3 \omega_4}=W_{mpqr}$ and the kinetic Eq.(\ref{eq:kin_eqn}-\ref{eq:Omega_mpqs}) recovers Eq.(3), and the RJ equilibrium Eq.(\ref{eq:n_rj_general}) recovers Eq.(4).


\section{Pure spatial case: Coupled second- and fourth-order moments equations}

\noindent
{\bf Moments equations:}
The starting point is the pure spatial model Eq.(5):
\begin{align}
i \partial_z {b}_m^S(z) = \beta_m {b}_m^S  -  \gamma_o \sum_{p,q,r} W_{mpqr}{b}_p^S {b}_q^{S*} {b}_r^S.
\label{eq:nls}
\end{align}
Here, we show that when nonresonant interactions dominate the dynamics, the wave turbulence kinetic equations do not exhibit an $H-$theorem of entropy growth. 

Considering the pure spatial dynamics of Eq.(\ref{eq:nls}), the coupled Eqs.(\ref{eq:n_m_1}-\ref{eq:J_m_1}) for the second and fourth order moments 
\begin{eqnarray*}
{n}_{m}^S(z)=\left< b_m^S b_m^{S*}\right>,  
\qquad
{J}_{mpqs}^S(z)=\left< b_m^{S*} b_p^S b_q^{S*} b_s^S\right>,
\end{eqnarray*}
reduce to
\begin{align}
& \partial_z n_m^S = + i\gamma_o  \sum_{p,q,s} W_{mpqs}  {J}_{mpqs}^S - i\gamma_o  \sum_{p,q,s} W_{mpqs}^* {J}_{mpqs}^{S*}  
\label{eq:n_{cl}}\\
& \partial_z {J}_{mpqs}^S  = i \Delta \beta_{mpqs} {J}_{mpqs}^S - 2 i \gamma_o  W_{mpqs}^* M_{mpqs}(n(z))  - 2 i \gamma_o R_{mpqs}(n(z))
\label{eq:J_0}
\end{align}
where $\Delta \beta_{mpqs} = {\beta}_m+{\beta}_q-{\beta}_p-{\beta}_s$, and
\begin{eqnarray*}
&&M_{mpqs} ({\bm n}^S) = n_p^S n_q^S n_s^S 
+n_m^S n_p^S n_s^S -n_m^S n_q^S n_s^S -n_m^S n_p^S n_q^S  ,\\
&&R_{mpqs}({\bm n}^S) =
 \delta_{q,s} \, U_{pm}({\bm n}) (n_p^S-n_m^S) n_q^S 
 +
 \delta_{q,p} \,
U_{sm}({\bm n}^S) (n_s^S-n_m^S) n_p^S 
 \delta_{m,s} \,
U_{pq}({\bm n}^S) (n_p^S-n_q^S) n_s^S 
 +
 \delta_{m,p} \,
U_{sq}({\bm n}^S) (n_s^S-n_q^S) n_m^S,
\\
&&U_{pq}({\bm n}^S) = \sum_{s} W_{pqss} n_s^S =\int u_p^*({\bf r}) u_q({\bf r}) \sum_s n_s^S |u_s({\bf r})|^2d{\bf r}.
\end{eqnarray*}
Eq.(\ref{eq:n_{cl}}-\ref{eq:J_0}) conserve the particle number and the total energy:
\begin{eqnarray*}
N=\sum_m n_m^S(z), \qquad H=\sum_m \beta_m n_m^S(z) - \frac{\gamma_o}{4} \sum_{mpqs} W_{mpqs} J_{mpqs}^S(z)+ W_{mpqs}^* J_{mpqs}^{S*}(z).
\end{eqnarray*}
In addition, Eq.(\ref{eq:n_{cl}}-\ref{eq:J_0}) are formally reversible, i.e., they are invariant under the change of variable 
$$
z \to -z, \quad J_{mpqs}^S \to J_{mpqs}^{S*}, \quad W_{mpqs} \to W^*_{mpqs}.
$$
Consequently Eq.(\ref{eq:n_{cl}}-\ref{eq:J_0}) do not exhibit an $H-$theorem of entropy growth.

\begin{figure}
\includegraphics[width=0.9\columnwidth]{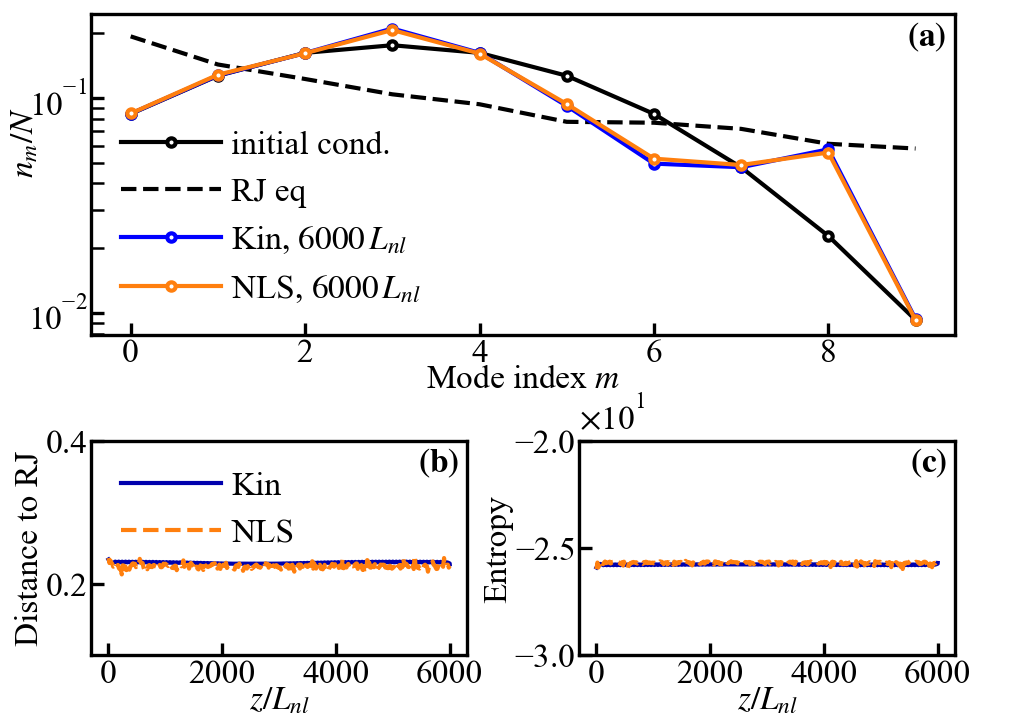}
\caption{
\baselineskip 10pt
{\bf Spatial case: Frozen thermalization.} 
The blue lines report the numerical simulation of the kinetic Eqs.(\ref{eq:n_{cl}}-\ref{eq:J_0}) showing the spectrum $n_m^S(z)$ at $z=6000 L_{\rm nl}$ (a), and corresponding evolutions of the distance to the RJ equilibrium ${\cal D}^S(z)$ (b), and the entropy (c).
The orange lines in (a)-(b)-(c) report the numerical simulations of Eq.(\ref{eq:nls}) governing the spatial modal amplitudes $b_m^S(z)$: Because of the large fluctuations, an average over 14 realizations has been taken by starting from the same initial spectrum (solid black line), with different realizations of the random phases.
A good agreement between the kinetic Eqs.(\ref{eq:n_{cl}}-\ref{eq:J_0}) and the spatial model Eq.(\ref{eq:nls}) is obtained without using adjustable parameters.
The dashed black line in (a) reports the expected RJ equilibrium spectrum.
The kinetic Eq.(\ref{eq:n_{cl}}-\ref{eq:J_0}) then explains the frozen thermalization of the purely spatial dynamics $b_m^S(z)$ [Eq.(\ref{eq:nls})] discussed  through the Fig.~3, as confirmed by the evolutions of the distance ${\cal D}^S(z)$ and the entropy in panels (b)-(c).
}
\label{fig:suppl_spatial_case} 
\end{figure}

\bigskip
\noindent
{\bf Numerical simulations of Eqs.(\ref{eq:n_{cl}}-\ref{eq:J_0}) explaining the frozen thermalization of the pure spatial dynamics:}\\
We have performed numerical simulations of the  coupled second-order and fourth-order moments Eqs.(\ref{eq:n_{cl}}-\ref{eq:J_0}), starting from the same initial condition as in Fig.~2. We compare the numerical results to those of the simulations of the purely spatial model Eq.(\ref{eq:nls}) reported in Fig.~3, where an average over 21 realizations was taken. The results are reported in Fig.~\ref{fig:suppl_spatial_case}.
We observe a quantitative agreement without using adjustable parameters. The kinetic Eqs.(\ref{eq:n_{cl}}-\ref{eq:J_0}) then explain the frozen process of thermalization observed in Fig.~3.

\bigskip
\noindent
{\bf Discussion:}
In order to derive the classical irreversible kinetic equation for the second-order moment, one needs to take the continuous limit of the discrete sums over the modes in Eq.(\ref{eq:n_{cl}}-\ref{eq:J_0}).
Before taking any limit, we note that the sums in Eq.(\ref{eq:n_{cl}}-\ref{eq:J_0}) involve three different types of resonances:

\smallskip
\noindent
{\bf (i)} The exact resonances correspond to combinations of the uples $\{m, p, q, s\}$ such that $\Delta \beta_{mpqs}=0$. 
These include in particular all trivial resonances involving only two spatial modes, i.e., ($m=p, q=s$), or ($m=s, q=p$).
The formal solution of Eq.(\ref{eq:J_0}) for $J_{mpqs}$ can be substituted in Eq.(\ref{eq:n_{cl}}), which gives for $\Delta \beta_{mpqs}=0$:
\begin{align}
\partial_z^2 n_m^S = 4  \gamma_o^2  \sum_{p,q,s}  \delta(  \Delta \beta_{mpqs} )  
 |W_{mpqs}|^2  M_{mpqs}(n)
+ 4 \gamma_o^2  \sum_{p,q,s}  \delta(  \Delta \beta_{mpqs} ) \Re[  W_{mpqs}  R_{mpqs}(n(z))   ].
\end{align}
This equation is formally reversible (second-order with respect to the `time' $z-$variable). Exact resonances occur in the case where the trapping potential $V(\br)$ exhibits a parabolic shape, due to the regular spacing of the eigenvalues.
In this case, the simulations evidence a quasi-reversible exchange of power among the modes, which tends to freeze thermalization process. This oscillatory behavior is related to the existence of Fermi-Pasta-Ulam recurrences, as discussed in Ref.\cite{biasi21} in the weakly nonlinear regime of the 2D NLS equation with a parabolic trapping potential. Note that in this latter case, RJ thermalization can be restored by introducing a weak disorder (random mode coupling) that breaks the coherent modal phase dynamics \cite{PRL19,Podivilov19}.


\smallskip
\noindent
{\bf (ii)} The non-resonant terms correspond to combinations of the uples $\{m, p, q, s\}$ such that $|\Delta \beta_{mpqs}| L_{\rm nl} \gg 1$. 
These terms are characterized by a rapid rotating phase of $J_{mpqs}(z)$ that averages to zero the evolutions of $n_m(z)$.

\smallskip
\noindent
{\bf (iii)} The quasi-resonant terms are the uples $\{m, p, q, s\}$ such that $|\Delta \beta_{mpqs}| L_{\rm nl} \lesssim 1$. 

\smallskip
\noindent
If quasi-resonances dominate over exact resonances, then the dynamics is irreversible \cite{nazarenko11}. This happens in the continuous limit, because there are many more quasi-resonances than exact resonances. In this limit, one recovers the classic kinetic equation that exhibits a $H-$theorem of entropy growth describing RJ thermalization. However, the continuous limit is not justified when one considers usual optical experiments with highly multimode step-index fibers.


\endwidetext

\section{Numerical methods}
\label{sec:num_meth}

The NLSE and NEE models are solved numerically by using a pseudo-spectral split-step method, with a Fourier truncated spectrum defined through a Galerkin truncation \cite{krstulovic11}. In order to accurately conserve the momentum $P_b$, we have implemented a dealiasing numerical procedure. In this way, $P_b$ and $N_b$  in Eqs.(\ref{eq:N_b_P_b}) are conserved at $10^{-7}$, and the Hamiltonian $H_b$ in Eqs.(\ref{eq:H_E_b}-\ref{eq:H_U_b}) at $10^{-4}$, throughout the simulation reported in Fig.~2 and Fig.~4, and in Figs.~\ref{fig:suppl_nee_global}-\ref{fig:suppl_RJ_loc} below.\\ 
In Fig.~2, we considered a step index waveguide that guides 10 modes  (core-cladding refractive index difference $\Delta n= 2\times 10^{-4}$), with a core radius of $\simeq 90\mu$m in which an ellipticity has been introduced to remove the mode degeneracies (with major-to-minor axis ratio 1.4), so as to be consistent with the derivation of the wave turbulence kinetic equation. 
The propagation constants $\{\beta_m\}$ and eigenfunctions $\{u_m(\br)\}$ have been computed at the central wavelength $\lambda_o=2.4 \mu$m (${\tilde \omega}_o=1.12\times 10^{3}$THz).
The propagation constants $\{\beta_m\}$ lie within the interval $\beta_0 = 19.6/L_{\rm nl}$ and $\beta_{9} \simeq 169/L_{\rm nl}$, so that the spatial mode dynamics evolves in the wealkly nonlinear regime, $\beta_p L_{\rm nl} \gg 1$, with $L_{\rm nl}=0.3$m. In all numerical simulations, the initial spectral widths are the same for all of the modes $\sim \exp(-\omega^2/\sigma_\omega^2)$, with $\sigma_\omega=1/\tau_0$ in Fig.~2.
We considered the anomalous dispersion regime ($\kappa_2<0$) with a defocusing nonlinearity ($\gamma<0$), to avoid the formation of temporal solitons, whose presence slows, or even freezes, the thermalization process. Note that the presence of temporal solitons can also be avoided in the focusing regime ($\gamma>0$) by considering the normal dispersion regime ($\kappa_2>0$). 
The simulations are realized in dimensionless units, and the corresponding typical parameters can be $\kappa_2=-0.1$ps$^2/$m with $L_{\rm nl}=0.3$m.\\
In Fig.~4, the investigation of the power-law scaling behavior $T^{loc}(z) \sim z^{-1/7}$ was computationally demanding. Accordingly, the number of modes, the nonlinear length, and the initial modal spectral width were decreased with respect to Fig.~2, so as to track the evolution toward the asymptotic behavior ($M=5, L_{\rm nl}=0.06$m, $\omega_c = 25/\tau_0, \sigma_\omega=0.4/\tau_0$). The initial modal amplitudes for the five modes were the same as in Fig.~2.

The data reported in Fig.~1  were generated using the same waveguide parameters and frequency grid (see the next section) as in the ST simulation shown in Fig.~2. 
With these parameters, we have reported in the inset of Fig.~1  the dispersion relations for the considered $M=10$ modes:
\begin{equation}
{\tilde \beta}_m(\omega)=\beta_m - \kappa_2 \omega^2=\beta_m - {\rm sgn}(\kappa_2) ( \omega \tau_0)^2/L_{\rm nl},
\end{equation} 
for $m=0,...,M-1$, and $L_{\rm nl}=0.3$m, $\tau_0=\sqrt{|\kappa_2| L_{\rm nl}}$.
In counting mode quadruplets in Fig.~1, we considered only non-degenerate four-mode resonances involving at least three distinct mode numbers, since trivial degenerate resonances involving only two modes do not contribute to thermalization.
In the ST case, the minimum $\Delta {\beta}^{ST}={\rm min}(\Delta {\tilde \beta}_{mpqr}^{1234})$ is taken over the ensemble of frequency combinations $\{\omega_1, \omega_2,\omega_3, \omega_4 \}$ under the constraint of momentum conservation, $\omega_1+\omega_3=\omega_2+\omega_4$.


To complete our numerical study based on NLSE in the main text, we present numerical simulations of the UPE (\ref{eq:upe1}-\ref{eq:upe2}) in the NEE approximation, see Eqs.(\ref{eq:limit_step_nee}-\ref{eq:Gamm_m_nee}). Fig.~\ref{fig:suppl_nee_global} reports the results of the NEE simulation. We observe a fast convergence toward the generalized RJ equilibrium distribution Eq.(\ref{eq:n_rj_general}). The main difference between the NLSE and NEE simulations is the presence of an asymmetric evolution of the time-spectrum, which essentially originates from the nontrivial spatio-temporal coupling of the dispersion relation ${\tilde \beta}(\omega)$ in Eq.(\ref{eq:limit_step_nee}), see Fig.~\ref{fig:suppl_nee_global}(d)-(e)-(f). This asymmetric spectral dynamics will be discussed in detail through the process of adiabatic cooling in Fig.~\ref{fig:suppl_RJ_loc}.

\begin{center}
\begin{figure}
\includegraphics[width=1\columnwidth]{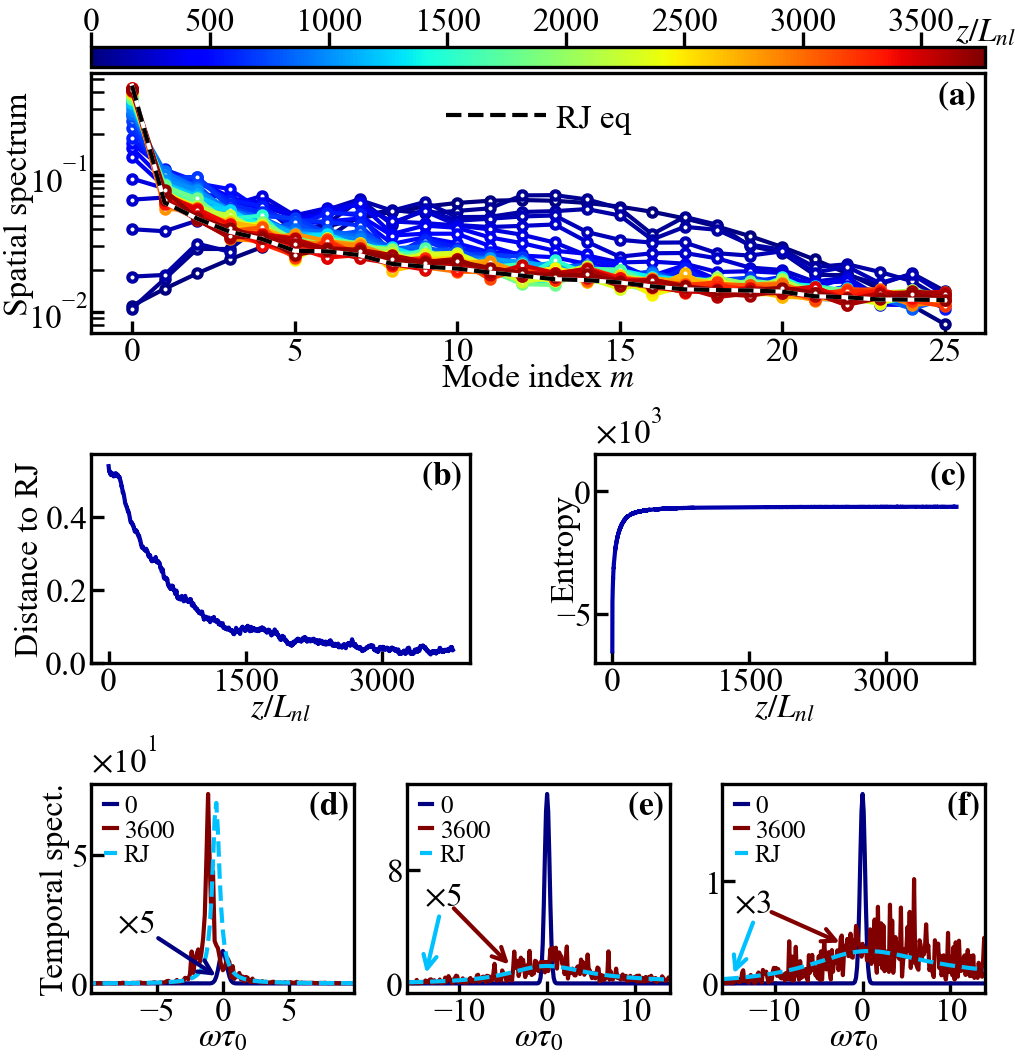}
\caption{
\baselineskip 10pt
{\bf Spatio-temporal thermalization: NEE simulation.} 
(a) Simulation of UPE (\ref{eq:upe1}-\ref{eq:upe2}) in the NEE approximation [see Eqs.(\ref{eq:limit_step_nee}-\ref{eq:Gamm_m_nee})]: Evolution of spatial modal occupation $N_m^{ST}/\sum_{m'}N_{m'}^{ST}$, showing the relaxation to the equilibrium RJ distribution Eq.(\ref{eq:n_rj_general}) (dashed black line).
(b) Evolution of the distance ${\cal D}^{ST}(z)$ to equilibrium, whose decrease to zero evidences ST thermalization. 
(c) This irreversible process is also characterized by a monotonous growth of entropy $S(z)$, as described by the $H-$theorem of the wave turbulence kinetic Eq.(\ref{eq:kin_eqn}).
Temporal spectrum $|b_m|^2(\omega,z)$  of the fundamental mode $m=0$ (d), intermediate mode $m=10$ (e), highest mode ($m=25$) (f), at $z=0$ (dark blue) and $z = 3600 L_{\rm nl}$ (red), showing thermalization to RJ spectra (dashed light blue).
Parameters: step-index waveguide supporting 26 modes, with anomalous dispersion and defocusing nonlinearity, $\tau_0=\sqrt{|\kappa_2| L_{\rm nl}}$, $L_{\rm nl}=0.06$m, ${\omega}_c=16/\tau_0$, $\kappa_2=-0.1$ps$^2/$m, ${\tilde \omega}_o=2.8\times 10^{3}$THz, $\sigma_\omega=0.4/\tau_0$, $\kappa_3=0.72\times 10^{-4}$ps$^3/$m.
}
\label{fig:suppl_nee_global} 
\end{figure}
\end{center}

\begin{center}
\begin{figure}
\includegraphics[width=1\columnwidth]{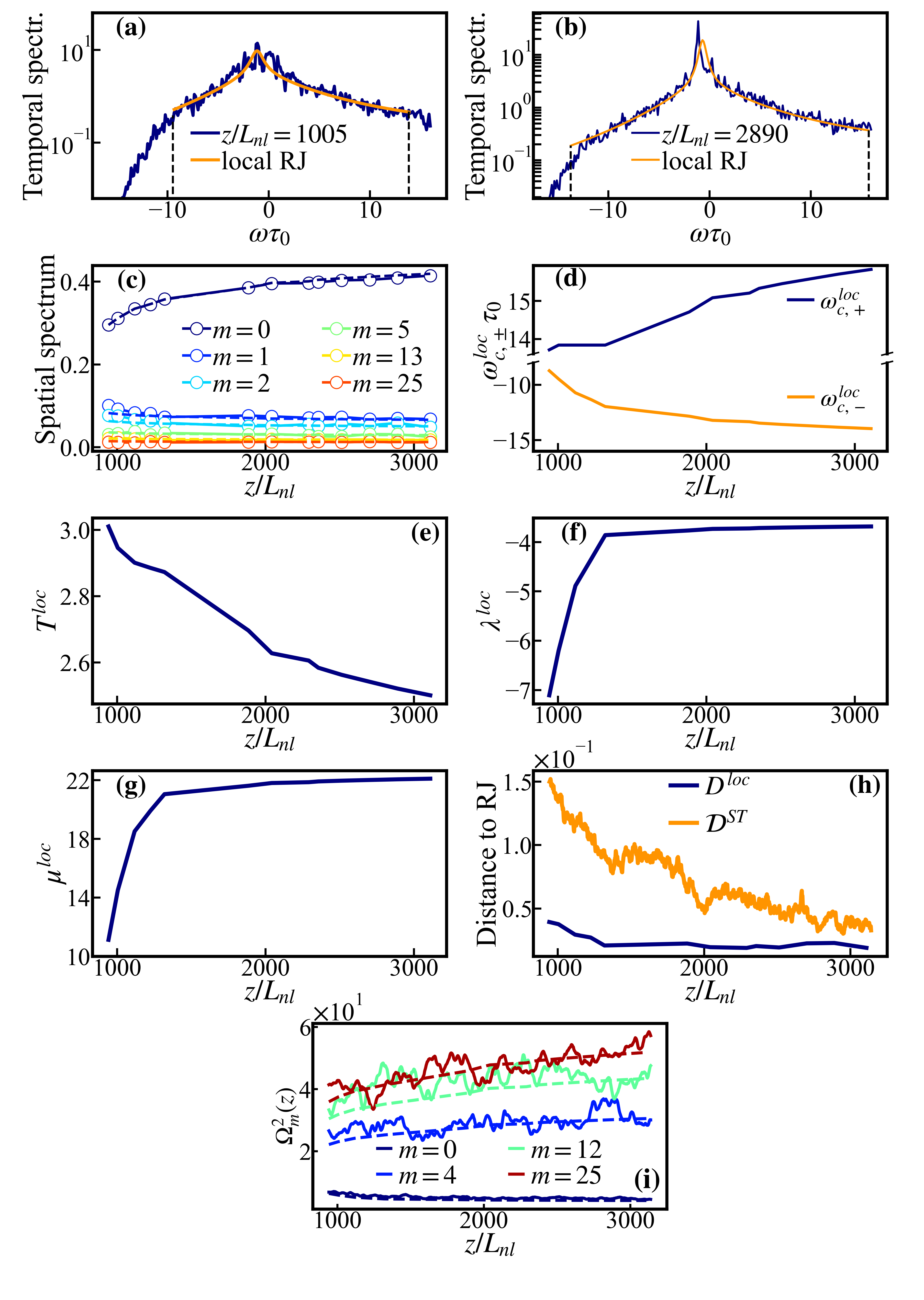}
\caption{
\baselineskip 10pt
{\bf Local-equilibrium route to ST thermalization for NEE simulation in Fig.~\ref{fig:suppl_nee_global}.} 
(a)-(b): Mode-integrated temporal spectrum of the field from the NEE simulation reported in Fig.~\ref{fig:suppl_nee_global} at $z=1005 L_{\rm nl}$ and $z=2890 L_{\rm nl}$ (blue line), and the corresponding local RJ equilibrium distribution computed over the reduced frequency window $[\omega_{c,-}^{loc},\omega_{c,+}^{loc}]$ (orange).
(c) Modal population $N_m^{loc}/\sum_{m'}N_{m'}^{loc}$ computed from the local RJ equilibrium (circles), and modal population $N_m^{ST}/\sum_{m'}N_{m'}^{ST}$ in the NEE simulation of Fig.~\ref{fig:suppl_nee_global} (continuous lines). 
Evolution during propagation of the local frequency cut-off ${\omega}_{c,\pm}^{loc}(z)$ (d), and local thermodynamic parameters characterizing the local RJ equilibrium: $T^{loc}(z)$ (e), $\lambda^{loc}(z)$ (f), and $\mu^{loc}(z)$ (g).
Panel (h) reports the local distance ${\cal D}^{loc}(z)$ computed over the reduced frequency interval $\omega \in [\omega_{c,-}^{loc}, \omega_{c,+}^{loc}]$ (blue line), and global distance ${\cal D}^{ST}(z)$ computed over the whole frequency window $[-\omega_{c}, \omega_{c}]$ (orange line). 
(i) Evolution during propagation of the normalized second-order moment $\Omega_m^2(z)$  of the time-spectrum of the field for different modes $m$, which are retrieved from the NEE simulation (solid lines), and from the local RJ equilibrium spectrum (dashed lines). Note that the fundamental mode $m=0$ shows spectral narrowing, signaling the onset of a ST beam cleaning process. 
}
\label{fig:suppl_RJ_loc} 
\end{figure}
\end{center}

\subsection{Global RJ equilibrium}

In this section, we discuss the computation of the {\it global RJ equilibrium} distribution throughout the time-frequency window available in the simulation $\omega \in [-\omega_c, \omega_c]$, as reported in Fg.~2  for NLSE, and in Fig.~\ref{fig:suppl_nee_global} for NEE. The case of {\it local RJ equilibrium} where the equilibrium distribution is computed over a truncated time-frequency window $\omega \in [\omega_{c,-}^{loc}, \omega_{c,+}^{loc}]$ (Fig.~4 for NLSE;  Fig.~\ref{fig:suppl_RJ_loc} for NEE), will be discussed in the next subsection.

The three parameters $(T,\lambda,\mu)$ involved in the RJ distribution (4)   are determined by the three conserved quantities $(E_n,P_n,N_n)$ in the original basis $\{n_m(\omega)\}$. Discretizing the frequency-domain integrals, we have
\begin{align}
E_n &= TM^{ST} \delta \omega+\lambda P_n +\mu N_n
\label{eq:eqstate}\\
\frac{P_n}{N_n} &= \frac{ \sum_{m,j} \frac{\omega_j }{\beta_m(1+\omega_j/{\tilde \omega}_o)^{-1} -\sum_{k \ge 2} \kappa_k \omega_j^k -\lambda \omega_j-\mu }
}{ \sum_{m,j} \frac{1}{\beta_m(1+\omega_j/{\tilde \omega}_o)^{-1} -\sum_{k \ge 2}\kappa_k \omega_j^k -\lambda \omega_j-\mu}}
\label{eq:PsN}\\
\frac{E_n}{N_n} &= \frac{ \sum_{m,j} \frac{\beta_m(1+\omega_j/{\tilde \omega}_o)^{-1} -\sum_{k \ge 2}\kappa_k \omega_j^k }{\beta_m(1+\omega_j/{\tilde \omega}_o)^{-1} -\sum_{k \ge 2}\kappa_k \omega_j^k -\lambda \omega_j-\mu } 
}{ \sum_{m,j} \frac{1}{\beta_m(1+\omega_j/{\tilde \omega}_o)^{-1} -\sum_{k \ge 2}\kappa_k \omega_j^k -\lambda \omega_j-\mu}}
\label{eq:EsN}
\end{align}  
where $M^{ST}=M Q$ is the total number of modes in the numerics ($M$ spatial modes, $Q$ temporal modes).  
The time-frequencies modes $\omega_j=j \delta \omega$ ($j=-Q/2-1,...,Q/2$, with $Q$ modes) result from the temporal grid used in the simulations (with periodic boundary conditions), where $\delta \omega=2\pi/{\cal T}$ is the frequency grid spacing and ${\cal T}$ the numerical time window. Accordingly, the frequency cut-off for the positive and negative frequencies are (approximately) the same, with 
\begin{equation}
\omega_{c}=\pi Q/{\cal T}.
\label{eq:w_pm}
\end{equation}
The two parameters $(\lambda,\mu)$ solutions of Eqs.(\ref{eq:PsN},\ref{eq:EsN}) are determined by an optimization algorithm. Then $T$ is obtained from Eq.(\ref{eq:eqstate}).
Note that, by increasing the number of modes $Q$ (while keeping fixed the frequency cut-off $\omega_c$), the discrete sums over the temporal modes $\omega_j$ can be converted to continuous integrals, and the parameters $(T,\lambda,\mu)$ solution of Eqs.(\ref{eq:eqstate}-\ref{eq:EsN}) converge to well-defined values, that is, {\it the RJ equilibrium distribution (4)  does not depend on the numerical discretization of the temporal grid}. The calculations reveal that the number of modes used in the simulations ($Q=256$) is sufficient to reach the continuous limit.
The restriction on the number of temporal modes results from the significant CPU time associated with the long propagation lengths $\sim 10^{6}\, \beta_{\rm max}^{-1}$ needed to reach thermal equilibrium ($\beta_{\rm max}^{-1}$ being the smallest propagation length scale).

\subsection{Local RJ equilibrium}
\label{subsec:Local_RJ_equilibrium}

As discussed through Fig.~4  for NLSE, and Fig.~\ref{fig:suppl_RJ_loc} for NEE, the optical field follows a local RJ equilibrium state during the propagation. The local thermodynamic parameters $(T^{loc}(z), \lambda^{loc}(z), \mu^{loc}(z))$ are computed over a reduced time frequency window $\omega \in [\omega_{c,-}^{loc}, \omega_{c,+}^{loc}]$. In the NLSE case, the spectrum is symmetric, so that $\omega_{c,-}^{loc}=- \omega_{c,+}^{loc}$. Here, we discuss the more general NEE case, where the evolution of the spectrum is highly asymmetric (see Fig.~\ref{fig:suppl_RJ_loc}), with  two distinct local frequency cut-off $\omega_{c,\pm}^{loc}$, for positive and negative frequencies:
\begin{equation}
-\omega_{c} \le \omega_{c,-}^{loc} < 0 < \omega_{c,+}^{loc} \le \omega_{c}.
\end{equation}
The fact that the field follows a local RJ equilibrium state during propagation is supported by the following numerical analysis:
(i) At some propagation length $z_0$, a local average of the optical field is computed over $z \in [z_0-\Delta z, z_0+\Delta z]$, with $\Delta z =3L_{\rm nl}$ (10 realizations).
(ii) At $z=z_0$, we compute the two local frequency cut-off, $\omega_{c,\pm}^{loc}(z_0)$, that minimize the distance ${\cal D}^{loc}(z_0)=\sum_m |N_m^{ST}(z_0) - N_m^{\rm RJ,loc}|  / \sum_m (N_m^{ST}(z_0) + N_m^{\rm RJ,loc})$, between the averaged optical field and the local RJ equilibrium, which is truncated over the compact support $[\omega_{c,-}^{loc}, \omega_{c,+}^{loc}]$, see Fig.~2(a)  for NLSE, or Fig.~\ref{fig:suppl_RJ_loc}(a)-(b) for NEE. This provides the local thermodynamic parameters: $(T^{loc}(z_0), \lambda^{loc}(z_0), \mu^{loc}(z_0))$ that characterize the local RJ equilibrium at $z=z_0$ over the reduced frequency window $[\omega_{c,-}^{loc},\omega_{c,+}^{loc}]$, see 
Fig.~4(c)  for NLSE, and Fig.~\ref{fig:suppl_RJ_loc}(d) for NEE. 

We report in Figs.~\ref{fig:suppl_RJ_loc}(a)-(b) the mode-integrated field temporal spectrum $\sum_m |b_m(\omega)|^2$ of the NEE simulation, and the corresponding local RJ equilibrium spectrum, at different propagation lengths. At variance with the negative frequency cutoff $\omega_{c,-}^{loc}$, the positive frequency cut-off increases rapidly and reaches the cut-off frequency of the spectral grid considered in the simulation, that is $\omega_{c,+}^{loc} \simeq \omega_c=16/\tau_0$, see Fig.~\ref{fig:suppl_RJ_loc}(d). The fact that the optical field follows a quasi-equilibrium path through propagation is reflected by the small value of the distance to the local RJ equilibrium, ${\cal D}^{loc}(z_0) < 0.05$ in Fig.~\ref{fig:suppl_RJ_loc}(h). For clarity, we have also plotted in Fig.~\ref{fig:suppl_RJ_loc}(h) the global distance ${\cal D}^{ST}(z)$ to the global RJ reported in Fig.~\ref{fig:suppl_nee_global}(b), which is computed with the frequency cut-off related to the numerical spectral grid, $\omega_{c}=16/\tau_0$. 

We stress in Fig.~\ref{fig:suppl_RJ_loc}(c) the remarkable agreement between the spatial modal population computed from the local RJ equilibrium $N_m^{loc}(z)$ (circles) and the actual modal population $N_m(z)$ in the NEE simulation (continuous lines).
{\it The fitting procedure is based on minimizing the distance ${\cal D}^{loc}(z)$, which means that nothing in the procedure explicitly enforces the agreement with the evolution of individual modes}, $N_m^{loc}(z)$ and $N_m^{ST}(z)$, shown in Fig.~\ref{fig:suppl_RJ_loc}(c). The same remark holds for the NLSE situations, see Fig.~4(b). 

We have also verified that the evolution of the spectral expansion $\omega_{c,\pm}^{loc}(z)$, and the associated temperature decay $T^{loc}(z)$, do not depend on the numerical frequency cut-off $\omega_c$, i.e., the local equilibrium state is not influenced by spectral boundaries imposed by the numerical grid. This of course requires that the numerical cut-off $\omega_c$ is taken large enough to accommodate the truncated classical distribution for the whole duration of the simulation, that is $|\omega_{c,\pm}^{loc}(z)|<\omega_c$ at all $z$. If this is not the case, the adiabatic cooling process is artificially stopped when $|\omega_{c,\pm}^{loc}(z)|$ hits the numerical cut-off $\omega_c$. In a fully quantum theory~\cite{chiocchetta16,davis01,blakie08}, an effective cut-off automatically appears at the frequency $\omega$ where the occupation $n_m(\omega,z)$ predicted by the RJ distribution falls below one.

Finally, we report in Fig.~\ref{fig:suppl_RJ_loc}(i) the evolution during propagation of the normalized second-order moment of the spectrum: 
\begin{align*}
\Omega_m^2(z) = \frac{\int {\omega}^2 |b_m({\omega},z)|^2 d{\omega}}{\int |b_m({\omega},z)|^2 d{\omega}}.
\end{align*}
The solid lines report the evolution of $\Omega_m^2(z)$ recovered from the NEE simulation (Fig.~\ref{fig:suppl_nee_global}), the dashed lines the evolution retrieved from the local RJ equilibrium spectra. It is interesting to note that the fundamental mode is the unique mode whose second-order moment decreases during propagation, indicating an incipient process of spatio-temporal beam cleaning.

\widetext
\section{Theory of adiabatic cooling: derivation of Eq.(6), $T^{loc} \sim z^{-1/7}$ }
\label{sec:ad_cool_th}

We have shown that the process of adiabatic cooling is characterized during propagation by a slow expansion of the spectral window  $[\omega_{c,-}^{loc},  \omega_{c,+}^{loc}]$ associated to the local RJ equilibrium, as illustrated in Fig.~4(c). In this section we provide a theoretical description of the spectral expansion and the associated process of adiabatic cooling Fig.~4(d), by using the wave turbulence kinetic equation derived above in Sec.~\ref{sec:deriv_KE}. We develop the theory in the NLSE approximation, see Eq.(2). Assuming that the initial spectrum is even, then the spectrum remains even during propagation, so that $\omega_{c,-}^{loc}(z) = - \omega_{c,+}^{loc}(z)$, at any $z> 0$. We recall that $\omega_{c}^{loc}(z)=\omega_{c,+}^{loc}(z)$ denotes the positive value of the local frequency cutoff.

We start from the kinetic Eq.(3).
The presence of the Dirac $\delta-$functions enables the computation of two integrals over the frequencies, so that the kinetic equation can be written in the following form:
\begin{align}
\partial_z n_m(\omega) = &4 \pi \gamma^2 \widetilde{\sum} \frac{|W_{mpqs}|^2}{2|\kappa_2|}  \int_0^\infty M_{mpqs}[{\bm n}](\omega ,\omega+ \Delta \omega_{mpqs}\nu y  ,\omega +\Delta \omega_{mpqs}\nu (y +\sigma_{mpqs}/y), \omega+\Delta \omega_{mpqs}\nu \sigma_{mpqs} /y) \frac{dy}{y}  
\nonumber\\
&+ 4 \pi \gamma^2 \sum_p
 \frac{|W_{mmpp}|^2}{|\kappa_2|}  \int [
 ( n_p(\omega)-n_m(\omega))   n_m(\omega') n_p(\omega')
 +n_m(\omega) n_p(\omega)( n_m(\omega')-n_p(\omega'))] \frac{d\omega'}{|\omega'-\omega|}
 \label{eq:kin1}
\end{align}
where $\widetilde{\sum}=\sum_{\nu\in \{-1,1\}}
\sum_{p,q,s, \Delta \beta_{mpqs} \neq 0}$, and
\begin{align*}
&M_{mpqs}  [{\bm n}](\omega_1,\omega_2,\omega_3,\omega_4) = 
n_m(\omega_1) n_p(\omega_2) n_q(\omega_3) n_s(\omega_4) \big( n_m^{-1}(\omega_1)+n_q^{-1}(\omega_3)-n_p^{-1}(\omega_2)-n_s(\omega_4)^{-1} \big)  ,\\
&\Delta \omega_{mpqs} = \sqrt{ |\Delta\beta_{mpqs}|/(2|\kappa_2|)} ,\\
&\sigma_{mpqs} = {\rm sgn}( \kappa_2 \Delta \beta_{mpqs}),\\
&\Delta\beta_{mpqs} =\beta_m -\beta_p +\beta_q -\beta_s .
\end{align*}
We recall that the spectra $n_m(\omega)$ are even for $z \ge 0$. It is then possible to show recursively that, at any distance $z$ the spectrum is essentially compactly supported in $[-\omega_c^{loc}(z),\omega_c^{loc}(z)]$ (or decays faster than any power for $|\omega| > \omega_c^{loc}(z)$).
If at some distance $z$ the spectrum is essentially compactly supported in $[-\omega_c^{loc}(z),\omega_c^{loc}(z)]$, then the dominant terms of the kinetic Eq.(\ref{eq:kin1}) 
(all terms that contain a factor at frequency $\omega$, i.e. all terms except the ones of the form Eq.(\ref{eq:kin2})) impose that the solution quickly takes the form of a local RJ distribution:
\begin{align*}
n_m(\omega,z) = n_m^{RJ,loc}(\omega,z) {\bf 1}_{[-\omega_c^{loc}(z),\omega_c^{loc}(z)]}(\omega), 
\end{align*}
with 
\begin{align}
n_{m}^{RJ,loc}(\omega,z) = \frac{T^{loc}(z)}{{\tilde \beta}_m(\omega)-\mu^{loc}(z)},
\label{eq:RJ_loc_nlse}
\end{align}
and ${\bf 1}_{[-\omega_c^{loc}(z),\omega_c^{loc}(z)]}(\omega)=1$ if and only if $\omega \in [-\omega_c^{loc}(z),\omega_c^{loc}(z)]$, and zero elsewhere. For a given $\omega_c^{loc}(z)$, the parameters $T^{loc}(z)$ and $\mu^{loc}(z)$ are determined by conservation of the total power (mass) $N_n$ and the energy $E_n$. We recall that the Lagrange multiplier $\lambda$ in the RJ equilibrium Eq.(\ref{eq:n_rj_general}) associated to the conservation of momentum $P_n$ vanishes, because the spectra are symmetric, i.e., $\lambda=P_n=0$. 
There are, additionally, terms in the kinetic equations that allow to increase the bandwidth $\omega_c^{loc}(z)$.
Indeed, 
if $\omega>\omega_c^{loc}(z)$, then 
\begin{align}
\partial_z n_m(\omega) = 4 \pi \gamma^2 
\sum_{p,q,s, \, \kappa_2 \Delta \beta_{mpqs}>0} \frac{|W_{mpqs}|^2}{2|\kappa_2|}  \int_0^\infty n_p(\omega- \Delta \omega_{mpqs} y )n_q(\omega - \Delta \omega_{mpqs} (y +1/y)) n_s(\omega- \Delta \omega_{mpqs} /y) \frac{dy}{y}  ,
\label{eq:kin2} 
\end{align}
and we have a similar equation for $\omega<-\omega_c^{loc}(z)$.
The spectrum width, therefore, slowly increases.
It is possible to study this slow process.
Let us assume  that  $\omega_c^{loc}(z)$ is large. Using the RJ form of the spectrum in $[-\omega_c^{loc}(z),\omega_c^{loc}(z)]$:
\begin{align}
\partial_z n_m(\omega) &= 
4 \pi \gamma^2 
\sum_{p,q,s, \, \kappa_2 \Delta \beta_{mpqs}>0} \frac{|W_{mpqs}|^2}{2|\kappa_2|}  \int_0^\infty 
 {\bf 1}_{ (\omega-\omega_c^{loc})/\Delta \omega_{mpqs} < y < \Delta \omega_{mpqs}  /(\omega-\omega_c^{loc})}
n_p^{RJ,loc}(\omega- \Delta \omega_{mpqs} y ) \nonumber\\
&\qquad \times n_q^{RJ,loc}(\omega - \Delta \omega_{mpqs} (y +1/y)) n_s^{RJ,loc}(\omega- \Delta \omega_{mpqs} /y) \frac{dy}{y} 
\label{eq:kin3} 
\end{align}
if $\omega \in [\omega_c^{loc}(z),\omega_c^{loc}(z)+\Delta \omega ]$, $\Delta \omega= \max_{m,p,q,s} \Delta \omega_{mpqs} =\sqrt{2}\sqrt{\beta_{M-1}-\beta_0}/\sqrt{|\kappa_2|}$, where we recall that $M$ is the number of modes.
According to Eq.(\ref{eq:eqstate}), the energy over the RJ frequency support reads $E_n \simeq 2 M T^{loc} \omega_c^{loc} + \mu^{loc} N_n \simeq 2 M T^{loc} \omega_c^{loc}$, so that the temperature scales as $T^{loc} \sim 1/\omega_c^{loc}$ by energy conservation. Therefore, the right-hand side of Eq.(\ref{eq:kin3}) is of order $\big(T^{loc}/(\omega_c^{loc})^{2}\big)^3 \sim (\omega_c^{loc})^{-9}$ and $n_m(\omega,z)$ for $\omega \in [\omega_c^{loc},\omega_c^{loc}+\Delta \omega]$ will reach values of the order of $n_p^{RJ,loc}(\omega) \sim T^{loc}/(\omega_c^{loc})^{2} \sim (\omega_c^{loc})^{-3}$ when $z$ increases by $\Delta z$ such that $\Delta z (\omega_c^{loc})^{-9}\sim (\omega_c^{loc})^{-3}$, i.e. when $\Delta z \sim (\omega_c^{loc})^{6}$. 
For such a $\Delta z$, the cut-off frequency $\omega_c^{loc}$ increases by $\Delta \omega \sim 1$, i.e. 
$\frac{\Delta \omega_c^{loc}}{\Delta z} \sim \frac{1}{(\omega_c^{loc})^6}$. 
Hence $\omega_c^{loc}$ satisfies 
$ \partial_z (\omega_c^{loc})^7 \sim 1$. In this way, we obtain 
\begin{equation}
\omega_c^{loc}(z) \sim z^{1/7}, \quad {\rm and} \quad  T^{loc}(z) \sim z^{-1/7}.
\label{eq:wc_T_z}
\end{equation}
We have reported in Fig.~4  numerical simulations of the NLSE that confirm the scaling law in Eq.(\ref{eq:wc_T_z}). The local temperature $T^{loc}(z)$, and the local frequency cut-off $\omega_c^{loc}(z)$ associated to the local RJ spectrum were reported in Fig.~4(c)-(d). These data  were fitted with a power law function, which is found in excellent agreement with the theoretical prediction Eq.(\ref{eq:wc_T_z}).

\endwidetext



\begin{thebibliography}{99}



\bibitem{carusotto22} I. Carusotto, Quantum Fluids of Light, in Encyclopedia of Condensed Matter Physics, 2nd edition. Available at arXiv:2211.10980.


\bibitem{bloch21}
J. Bloch, I. Carusotto, M. Wouters, 
Nonequilibrium Bose–Einstein condensation in photonic systems, 
Nature Phys. Rev. {\bf 4}, 470 (2022).

\bibitem{glorieux25}
Q. Glorieux, C. Piekarski, Q. Schibler, T. Aladjidi, M. Baker-Rasooli,
Paraxial fluids of light,
Advances In Atomic, Molecular, and Optical Physics {\bf 74}, 157-241 (2025).


\bibitem{PRL20}
K. Baudin, A. Fusaro, K. Krupa, J. Garnier, S. Rica, G. Millot, A. Picozzi,
Classical Rayleigh-Jeans condensation of light waves: 
Observation and thermodynamic characterization,
Phys. Rev. Lett. {\bf 125}, 244101 (2020).


\bibitem{pourbeyram22}
H. Pourbeyram, P. Sidorenko, F. Wu, L. Wright, D. Christodoulides, F. Wise, Direct
observations of thermalization to a Rayleigh–Jeans distribution in multimode
optical fibres, Nature Physics {\bf 18}, 685 (2022).


\bibitem{mangini22}
F. Mangini, M. Gervaziev, M. Ferraro, D. S. Kharenko, M. Zitelli, Y. Sun, V. Couderc, E.V. Podivilov, S.A. Babin, and S. Wabnitz,
Statistical mechanics of beam self-cleaning in GRIN multimode optical fibers, 
Opt. Exp. {\bf 30}, 10850 (2022).

\bibitem{wright16}
L.G. Wright, Z. Liu, D.A. Nolan, M.-J. Li, D.N. Christodoulides, F.W. Wise,
Self-organized instability in graded-index multimode fibres,
Nature Photonics {\bf 10}, 771 (2016).

\bibitem{krupa17}
K. Krupa, A. Tonello, B.M. Shalaby, M. Fabert, A. Barth\'el\'emy, G. Millot, S. Wabnitz, V. Couderc,
Spatial beam self-cleaning in multimode fibres,
Nature Photonics {\bf 11}, 237 (2017).

\bibitem{ferraro23} For a review, see, M. Ferraro, F. Mangini, M. Zitelli, S. Wabnitz,
On spatial beam self-cleaning from the perspective of optical wave thermalization in multimode graded-index fibers,
Advances in Physics: X {\bf 8}, 2228018 (2023).




\bibitem{christodoulides19}
F.O. Wu, A.U. Hassan, D.N. Christodoulides,
Thermodynamic theory of highly multimoded nonlinear optical systems,
Nature Photonics {\bf 13}, 776 (2019).


\bibitem{PRA19}
J. Garnier, A. Fusaro, K. Baudin, C. Michel, K. Krupa, G. Millot, A. Picozzi, 
Wave condensation with weak disorder versus beam self-cleaning in multimode fibers,
Phys. Rev. A {\bf 100}, 053835 (2019).


\bibitem{kottos20}
A. Ramos, L. Fern\'andez-Alc\'azar, T. Kottos, B. Shapiro,
Optical Phase Transitions in Photonic Networks: a Spin-System Formulation,
Phys. Rev. X {\bf 10}, 031024 (2020).

\bibitem{makris22}
F.O. Wu, Q. Zhong, H. Ren, P.S. Jung, K.G. Makris, D.N. Christodoulides,
Thermalization of Light's orbital angular momentum in nonlinear multimode waveguide systems, 
Phys. Rev. Lett. {\bf 128}, 123901 (2022).



\bibitem{PRL23}
K. Baudin, J. Garnier, A. Fusaro, N. Berti, C. Michel, K. Krupa, G. Millot, A.
Picozzi, 
Observation of light thermalization to negative temperature Rayleigh-Jeans equilibrium states in multimode optical fibers, 
Phys. Rev. Lett. {\bf 130}, 063801 (2023).

\bibitem{ferraro25}
M. Ferraro, F. Mangini, K. Stef\'anska, W.A. Gemechu, F. Frezza, V. Couderc, M. Gervaziev, D. Kharenko, S. Babin, and S. Wabnitz,
Negative absolute temperature attractor in a dense photon gas,
arXiv:2505.21163

\bibitem{ferraro24} 
M. Ferraro, F. Mangini, F.O. Wu, M. Zitelli, D.N. Christodoulides, S. Wabnitz,
Calorimetry of photon gases in nonlinear multimode optical fibers, 
Phys. Rev. X {\bf 14}, 021020 (2024). 

\bibitem{kirsch25} 
M.S. Kirsch, G.G. Pyrialakos, R. Altenkirch, M.A. Selim, J. Beck, T.A.W. Wolterink, H. Ren, P.S. Jung, M. Khajavikhan, A. Szameit, M. Heinrich, D.N. Christodoulides, 
Observation of Joule--Thomson photon-gas expansion, 
Nat. Phys. {\bf 21}, 214–220 (2025).

\bibitem{podivilov22}
E.V. Podivilov, F. Mangini, O.S. Sidelnikov, M. Ferraro, M. Gervaziev, D.S.
Kharenko, M. Zitelli, M.P. Fedoruk, S.A. Babin, S. Wabnitz, 
Thermalization of orbital angular momentum beams in multimode optical fibers, 
Phys. Rev. Lett. {\bf 128}, 243901 (2022).

\bibitem{PRL22}
N. Berti, K. Baudin, A. Fusaro, G. Millot, A. Picozzi, J. Garnier, 
Interplay of thermalization and strong disorder: Wave turbulence theory, numerical simulations, and experiments in multimode optical fibers, 
Phys. Rev. Lett. {\bf 129}, 063901 (2022).

\bibitem{lu24}
M. Lian, Y. Geng, Y.-J. Chen, Y. Chen, J.-T. L\"u,
Coupled Thermal and Power Transport of Optical Waveguide Arrays: Photonic Wiedemann-Franz Law and Rectification Effect,
Phys. Rev. Lett. {\bf 133}, 116303 (2024).

\bibitem{kottos24}
A. Kurnosov, L. J. Fernández-Alcázar, A. Ramos, B. Shapiro,  T. Kottos,
Optical Kinetic Theory of Nonlinear Multimode Photonic Networks,
Phys. Rev. Lett. {\bf 132}, 193802 (2024).


















\bibitem{larre2015propagation}
P.-E. Larr{\'e}, I. Carusotto,
Propagation of a quantum fluid of light in a cavityless nonlinear optical medium: General theory and response to quantum quenches,
Physical Review A {\bf 92}, 043802 (2015).

\bibitem{glorieux22}
J. Steinhauer, M. Abuzarli, T. Aladjidi, T. Bienaim\'e, C. Piekarski, W. Liu, E. Giacobino, A. Bramati, Q. Glorieux, Analogue cosmological particle creation in an ultracold quantum fluid of light,
Nature Comm. {\bf 13}, 2890 (2022). 





\bibitem{agrawal}
G. Agrawal, {\it Nonlinear Fiber Optics} 
(Academic, New York, 6th ed., 2019).


\bibitem{wright22}
L.G. Wright, F.O. Wu, D.N. Christodoulides, F.W. Wise, 
Physics of highly multimode nonlinear optical systems,
Nature Physics {\bf 18}, 1018 (2022).




\bibitem{sun24}
Y. Sun, P. Parra-Rivas, G.P. Agrawal, T. Hansson, C. Antonelli, A. Mecozzi, F. Mangini, S. Wabnitz,
Multimode solitons in optical fibers: a review,
Photon. Research {\bf 12}, 2581 (2024).




\bibitem{kibler21}
B. Kibler, P. B\'ejot,
Discretized Conical Waves in Multimode Optical Fibers,
Phys. Rev. Lett. {\bf 126}, 023902 (2021).





\bibitem{krupa16}
K. Krupa, A. Tonello, A. Barth\'el\'emy, V. Couderc, B.M. Shalaby, A. Bendahmane, G. Millot, S. Wabnitz, 
Observation of geometric parametric instability induced by the periodic spatial self-imaging of multimode waves,
Phys. Rev. Lett. {\bf 116}, 183901 (2016).





\bibitem{gentyNC22}
Z. Eslami, L. Salmela, A. Filipkowski, D. Pysz, M. Klimczak, R. Buczynski, J.M. Dudley, G. Genty,
Two octave supercontinuum generation in a non-silica graded-index multimode fiber, 
Nature Comm. {\bf 13}, 2126 (2022). 

\bibitem{PR14}
A. Picozzi, J. Garnier, T. Hansson, P. Suret, S. Randoux, G. Millot,  D.N. Christodoulides, 
Optical wave turbulence: Toward a unified nonequilibrium thermodynamic formulation of statistical nonlinear optics,
Physics Reports {\bf 542}, 1-132 (2014).


\bibitem{chiocchetta16} 
A. Chiocchetta, P.E. Larr\'e, I. Carusotto, 
Thermalization and Bose-Einstein condensation of quantum light in bulk nonlinear media,
Europhys. Lett. {\bf 115}, 24002 (2016).

\bibitem{Mandonnet2000} Y. Castin, R. Dum, E. Mandonnet, A. Minguzzi and I. Carusotto,
Coherence properties of a continuous atom laser, 
Journal of Modern Optics {\bf 47}, 2671 (2000).

\bibitem{DGO_th} 
E.  Mandonnet, A.  Minguzzi, R. Dum, I. Carusotto, Y. Castin, J. Dalibard,  
Evaporative cooling of an atomic beam, 
Eur. Phys. J. D {\bf 10}, 9 (2000).

\bibitem{DGO_exp}  
T. Lahaye, Z. Wang, G. Reinaudi, S.P. Rath, J. Dalibard, D. Gu\'ery-Odelin,
Evaporative cooling of a guided rubidium atomic beam, 
Phys. Rev. A {\bf 72}, 033411 (2005).

\bibitem{Raithel} 
S.E. Olson, R.R. Mhaskar, G. Raithel, 
Continuous propagation and energy filtering of a cold atomic beam in a long high-gradient magnetic atom guide, 
Phys. Rev. A {\bf 73}, 033622 (2006).

\bibitem{Schreck} 
C.-C. Chen, S. Bennetts, R.G. Escudero, B. Pasquiou, F. Schreck, 
Continuous guided strontium beam with high phase-space density, 
Phys. Rev. Appl. {\bf 12}, 044014 (2019).


\bibitem{davis01} 
M.J. Davis, S.A. Morgan, K. Burnett,
Simulations of Bose Fields at Finite Temperature,
Phys. Rev. Lett. {\bf 87}, 160402 (2001).

\bibitem{PRL05} 
C. Connaughton, C. Josserand, A. Picozzi, Y. Pomeau, S. Rica, 
Condensation of classical nonlinear waves,
Phys. Rev. Lett. {\bf 95}, 263901 (2005).

\bibitem{blakie08} 
P.B. Blakie, A.S. Bradley, M.J. Davis, R.J. Ballagh, C.W. Gardiner, 
Dynamics and statistical mechanics of ultra-cold Bose gases using c-field techniques, 
Advance in Physics {\bf 57}, 363-455 (2008).



\bibitem{krstulovic11}
G. Krstulovic, M. Brachet,
Energy cascade with small-scale thermalization, counterflow metastability, and anomalous velocity of vortex rings in Fourier truncated Gross-Pitaevskii equation,
Phys. Rev. E {\bf 83}, 066311 (2011).

\bibitem{nazarenko11} 
S. Nazarenko, 
{\it Wave Turbulence} (Springer, Lectures Notes in Physics, 2011).





\bibitem{lvov10} 
V.S. L'vov, S.V. Nazarenko, 
Discrete and mesoscopic regimes of finite-size wave turbulence, 
Phys. Rev. E {\bf 82}, 056322 (2010).



\bibitem{onorato23}
M. Onorato, Y.V. Lvov, G. Dematteis, S. Chibbaro,
Wave Turbulence and thermalization in one-dimensional chains,
Physics Reports {\bf 1040}, 1 (2023).

\bibitem{nazarenko23}
Y. Zhu, B. Semisalov, G. Krstulovic, S. Nazarenko,
Self-similar evolution of wave turbulence in Gross-Pitaevskii system,
Phys. Rev. E {\bf 108}, 064207 (2023).

\bibitem{Gasenzer23}
For a review, see A.N. Mikheev, I. Siovitz, T. Gasenzer, Universal dynamics
and non-thermal fixed points in quantum fluids far from
equilibrium, Eur. Phys. J. Spec. Top. 232:3393-3415 (2023).




\bibitem{erne18}
S. Erne, R. B\"ucker, T. Gasenzer, J. Berges, J. Schmiedmayer, 
Universal dynamics in an isolated one-dimensional Bose gas far from equilibrium,
Nature {\bf 563}, 225 (2018).

\bibitem{prufer18}
M. Pr\"ufer, P. Kunkel, H. Strobel, S. Lannig, D. Linnemann, C.-M. Schmied, J. Berges, T. Gasenzer, and M.K. Oberthaler, 
Observation of universal dynamics in a spinor Bose gas far from equilibrium, 
Nature  {\bf 563}, 217 (2018).




\bibitem{moreno25}
M.A. Moreno-Armijos, A.R. Fritsch, A.D. Garc\'{\i}a-Orozco, S. Sab, G. Telles, Y. Zhu, L. Madeira, S. Nazarenko, V.I. Yukalov, V.S. Bagnato,
Observation of relaxation stages in a nonequilibrium closed quantum system: Decaying turbulence in a trapped superfluid, Phys. Rev. Lett. {\bf 134}, 023401 (2025).

\bibitem{gazo25}
M. Gazo, A. Karailiev, T. Satoor, C. Eigen, M. Galka, Z. Hadzibabic,
Universal coarsening in a homogeneous two-dimensional Bose gas,
Science  {\bf 389}, 802 (2025).


\bibitem{kolesik04}
M. Kolesik, J.V. Moloney, 
Nonlinear optical pulse propagation simulation: From Maxwell's to unidirectional equations,
Phys. Rev. E {\bf 70}, 036604 (2004).

\bibitem{biasi21}
A. Biasi, O. Evnin, and B.A. Malomed, 
Fermi-Pasta-Ulam phenomena and persistent breathers in the harmonic trap,
Phys. Rev. E {\bf 104}, 034210 (2021).

\bibitem{PRL19}
A. Fusaro, J. Garnier, K. Krupa, G. Millot, A. Picozzi, 
Dramatic acceleration of wave condensation mediated by disorder in multimode fibers, 
Phys. Rev. Lett. {\bf 122}, 123902 (2019).

\bibitem{Podivilov19}
E. Podivilov, D. Kharenko, V. Gonta, K. Krupa, O.S. Sidelnikov, S. Turitsyn, M.P. Fedoruk, S.A. Babin, S. Wabnitz, 
Hydrodynamic 2D turbulence and spatial beam condensation in multimode optical fibers, 
Phys. Rev. Lett. {\bf 122}, 103902 (2019).

\bibitem{kolesik02}
M. Kolesik, J.V. Moloney, and M. Mlejnek,
Unidirectional Optical Pulse Propagation Equation,
Phys. Rev. Lett. {\bf 89}, 283902 (2002).


\bibitem{couairon07} 
A. Couairon and A. Mysyrowicz, 
Femtosecond filamentation in transparent media, 
Physics Reports {\bf 441}, 47 (2007).

\bibitem{kolesik11}
A. Couairon, E. Brambilla, T. Corti, D. Majus, O. de J. Ramirez-Gongora, and M. Kolesik,
Practitioner's guide to laser pulse propagation models and simulation,
Eur. Phys. J. Special Topics {\bf 199}, 5-76 (2011).


\bibitem{moloney16} A. D. Bandrauk, E. Lorin, and J.V. Moloney, 
Laser Filamentation, Mathematical Methods and Models (Springer, 2016).




\bibitem{bejot19}
P. B\'ejot,
Multimodal unidirectional pulse propagation equation,
Phys. Rev. E {\bf 99}, 032217 (2019).


\bibitem{zakharov92}
V.E. Zakharov, V.S. L'vov, G. Falkovich, 
{\it Kolmogorov Spectra of Turbulence I} (Springer, Berlin, 1992).

\bibitem{Newell01} 
A.C. Newell, S. Nazarenko, L. Biven, 
Wave turbulence and intermittency, 
Physica D  {\bf 152}, 520 (2001).

\bibitem{Newell-Rumpf} 
A.C. Newell, B. Rumpf, 
Wave Turbulence,
Annu. Rev. Fluid Mech. {\bf 43}, 59 (2011).

\bibitem{shrira-nazarenko13}
S.K. Turitsyn, S.A. Babin, E.G. Turitsyna, G. E. Falkovich, E. Podivilov, D. Churkin, 
Optical Wave Turbulence, in {\it Advances in Wave Turbulence},  World Scientific Series on Nonlinear Science Series A, Vol. 83, edited by V.I. Shrira and S. Nazarenko (World Scientific, Singapore, 2013).

\bibitem{babin07}
S.A. Babin, D.V. Churkin, A.E. Ismagulov, S.I. Kablukov, E.V. Podivilov, 
Four-wave-mixing-induced turbulent spectral broadening in a long Raman fiber laser, 
J. Opt. Soc. Am. B {\bf 24}, 1729-1738 (2007).

\bibitem{churkin15}
D. Churkin, I. Kolokolov, E. Podivilov, I. Vatnik, S. Vergeles, I. Terekhov, V. Lebedev, G. Falkovich, M. Nikulin, S. Babin, and S.K. Turitsyn, 
Wave kinetics of a random fibre laser, 
Nature Comm. {\bf 2}, 6214 (2015).

\bibitem{laurie12}
J. Laurie, U. Bortolozzo, S. Nazarenko, S. Residori, 
One-dimensional optical wave turbulence: experiment and theory, 
Physics Reports {\bf 514}, 121-175 (2012).













\bibitem{wang20}
Z. Wang, W. Fu, Y. Zhang, H. Zhao,
Wave-turbulence origin of the instability of Anderson localization against many-body interactions,
Phys. Rev. Lett. {\bf 124}, 186401 (2020).

\bibitem{cherroret21}
N. Cherroret, T. Scoquart, D. Delande,
Coherent multiple scattering of out-of-equilibrium interacting Bose gases,
Annals of Physics {\bf 435}, 168543 (2021). 

\bibitem{kottos23}
A.Y. Ramos, C. Shi, L.J. Fernandez-Alcazar, D.N. Christodoulides, T. Kottos, 
Theory of localization-hindered thermalization in nonlinear multimode photonics,
Communications Physics {\bf 6}, 189 (2023).

\bibitem{shepelyansky23} 
K.M. Frahm and D.L. Shepelyansky,
Nonlinear Perturbation of Random Matrix Theory,
Phys. Rev. Lett. {\bf 131}, 077201 (2023).







































































































\end{thebibliography}

\end{document}